\documentclass[aps,pre,twocolumn]{revtex4}
\usepackage[dvips]{graphics}
\usepackage{graphicx}
\usepackage{amsfonts}
\usepackage{amssymb}
\usepackage{amsmath}
\usepackage{epsfig}

\def\rootfig{./figures/}
\def\sech{{\rm sech}}

\begin{document}

\title{Solitons Riding on Solitons:\\[1.0ex]
Hypersolitons and the Quantum Newton's Cradle}

\author{%
Manjun Ma$^1$,
R.\ Navarro$^2$, and
R.\ Carretero-Gonz\'{a}lez$^3$
}

\affiliation{
$^1$Department of Mathematics, School of Science, Zhejiang Sci-Tech University, Hangzhou, Zhejiang, 310018, China; mmj@cjlu.edu.cn\\
$^2$University of California San Diego,
Department of Mechanical and Aerospace Engineering,
9500 Gilman Drive-MC 0411, La Jolla, CA 92093-0411, USA\\
$^3$Nonlinear Dynamical Systems Group%
\footnote{URL: {\tt http://nlds.sdsu.edu/}},
Department of Mathematics and Statistics,
and Computational Science Research Center%
\footnote{URL: {\tt http://www.csrc.sdsu.edu/}},
San Diego State University, San Diego CA, 92182-7720, USA }

\begin{abstract}
The reduced dynamics for dark and bright soliton chains in the one-dimensional
nonlinear Schr\"odinger equation is used to study the behavior of collective
compression waves.
For appropriate conditions,
the reduced dynamics derived from perturbation and variational
techniques allows to describe a chain of dark or bright solitons
as a chain of effective masses connected by nonlinear springs
taking the form of a Toda lattice model on the soliton's positions.
In turn, the Toda lattice possesses exact solitary travelling
compression wave solutions corresponding to travelling
compression waves in the original soliton chain.
We coin the term hypersoliton to describe such solitary waves
riding on a chain of solitons.
We corroborate our analytical results with direct numerical simulations
of the nonlinear Schr\"odinger equation. It is observed that in the
case of dark soliton chains, the formulated reduction dynamics provides an
accurate description for the robust evolution of travelling compression
waves. In contrast, bright soliton chains do not have such stable
propagating solutions due to the desynchronization of the mutual
phases between consecutive solitons during evolution.
Finally, as an application to Bose-Einstein condensates
trapped in a standard external harmonic trapping potential,
we study the case of finite dark soliton chain confined at the
center of the trap.
We find that when the central chain is hit by a dark soliton initiated
at the edge of the external trap, the energy is transferred through
the chain as a hypersoliton. When the hypersoliton reaches the
end of the central chain, a dark soliton is ejected away from the
center of the trap and, as it returns from its excursion up the
trap, hits the central chain producing again a hypersoliton.
This periodic evolution is the equivalent of
the classical Newton's cradle.
\end{abstract}

\date{\today}


\maketitle

\section{Introduction}

Bose-Einstein condensation occurs when a dilute gas of bosonic atoms
is cooled below a critical temperature where a considerable
fraction of the atoms occupy the same quantum state
according to Bose-Einstein statistics.
Bose-Einstein condensates (BECs) were first theorized by Bose and Einstein
in the 1920s~\cite{Bose_1924} but not experimentally realized until
1995~\cite{Cornell_1995,Ketterle_1995} for which the 2001 Nobel Prize in
Physics was awarded~\cite{Nobel_2001}.
Typically, rubidium or sodium atoms are used and are
cooled to nanokelvin temperatures using a combination of laser an
evaporative cooling. The condensate is held in position by a
combination of magnetic and optical traps.
For sufficiently low temperatures, the mean field dynamics of
BECs in a quasi-one-dimensional (1D) 
trap can be accurately described by the so-called
Gross-Pitaevskii (GP) equation that is a variant of the nonlinear
Schr\"odinger (NLS) equation incorporating the external
trapping potential~\cite{BEC_BOOK}.
By appropriately adimensionalizing time, length and energy
(see Ref.~\cite{BEC_BOOK} for details), it is possible to cast the
1D GP equation as
\begin{equation}
i\,u_t=-\frac{1}{2} u_{xx} + g\, |u|^2 u + V_{\rm MT}\, u,
\label{GPE}
\end{equation}
where the rescaled condensate wavefunction is given by $u(x,t)$,
$V_{\rm MT}(x)$ is the effective 1D (magnetic) trapping potential
confining the BEC and $g=\pm 1$ indicates whether the atoms
have an attractive ($g=-1$) or repulsive ($g=+1$) scattering
length.
This 1D reduction of the system is achieved by the so-called
cigar-shaped external trapping potential for which two transverse
directions are tightly confining (such that, effectively, only
the ground state along these direction is possible) while the
longitudinal (in our case $x$) direction is loosely trapped allowing
for the dynamics of Eq.~(\ref{GPE}) to evolve along this direction.

Since the experimental realization of Bose-Einstein condensation,
the study of this new form of matter has
been the focus of intensive theoretical and experimental
efforts~\cite{pethick,stringari,BEC_BOOK}.
BECs continue to be a testbed for accessing quantum mechanics at a
macroscopic level allowing for direct observation of matter-wave
solitons~\cite{Dalfovo_99,BEC_BOOK}.
Under strong transverse confinement in two spatial directions, a BEC
can be rendered effectively quasi-1D~\cite{BEC_BOOK}.
In this case, depending on the sign of the scattering length between
the BEC entities (usually alkali atoms), it is possible to observe
bright~\cite{bright1,bright2,bright3} (for attractive interactions) and
dark~\cite{han1,nist,dutton,chap01:dark} (for repulsive interactions) solitons.
In the present work, we are interested in studying the collective
dynamics of chains of these 1D solitons and, specifically, the
possibility of stable solutions that coherently propagate
compression waves along the soliton chain.

Solitons are ubiquitous nonlinear waves that occur in a wide range
of physical systems such as plasmas, molecular chains, optical
fibers, and long water waves~\cite{Scott:book}.
In many physically relevant setups solitons are extremely robust (with
respect to parametric perturbations) and stable (with respect to
configuration perturbations). They can interact elastically with other
solitons, travel long distances, and travel through inhomogeneities
with minimal deformation and dispersion. This striking stability
relies on the balance between dispersion and nonlinearity.
%
%
For instance, in the absence of external trapping ($V_{\rm MT}=0$)
and for the case of an attractive condensate ($g=-1$),
the homogeneous background GP (\ref{GPE}) accepts exact {\em bright}
soliton (BS) solutions of the form~\cite{BEC_BOOK,nonlineairy_review}
\begin{equation}
u_{\rm bs} = a\, {\sech}(a (x -\xi(t)))\, e^{i\, (vx + \phi_{\rm bs}(t))},
\label{eq:BS}
\end{equation}
where $a$ is the amplitude (and inverse width) of the soliton,
$\xi(t)= v t + \xi_0$ is its position ($\xi_0$ being its
initial location) and its phase is given by
$\phi_{\rm bs}(t)=(a^2-v^2)t/2 + \phi_0$ ($\phi_0$ being its
initial phase).
On the other hand, in the case of a repulsive condensate,
for a homogeneous and stationary background density, the
GP equation (\ref{GPE}) accepts exact {\em dark} soliton (DS)
solutions of the form~\cite{DJF_REVIEW,DARK_BOOK}
\begin{equation}
u_{\rm ds} = \sqrt{n_0}\left[B\, {\tanh}(\sqrt{n_0}B\, (x -\xi(t)))+ iA\right] \, e^{i\phi_{\rm ds}(t)},
\label{eq:DS}
\end{equation}
where $n_0$ is the density of the constant background that supports
the DS, $\xi(t)$ is its position as defined above for a BS and its
phase is given by $\phi_{\rm ds}(t)=- n_0 t + \phi_0$ ($\phi_0$ being
its initial phase). The parameters of the DS are related by the
following expressions: $A^2+B^2=1$ and $v=A\sqrt{n_0}$.

In this work we study the collective dynamics of chains of interaction
bright and dark solitons. The manuscript is organized as follows.
The next section is devoted to describing the dynamical reduction
where the evolution of a chain of well separated and nearly identical solitons
can be reduced, for both BSs and DSs, to a chain of effective particles
connected with nonlinear springs modelled by the fundamental Toda lattice
on the soliton's positions.
Section~\ref{sec:TL}
is devoted to constructing appropriate initial conditions for the original
GP equation to support Toda lattice solitons riding on chains of BSs and DSs,
a.k.a~{\em hypersolitons}.
We present numerical results from direct integration of the GP equation
which very closely match the solutions of the corresponding Toda lattice
prediction.
We describe the robustness of the constructed hypersoliton solutions
and present some typical collision scenarios.
Also, in this section,
motivated by the presence of harmonic trapping in typical BEC
experiments, we study the effects of considering a finite
soliton chain that is confined at the bottom of the external potential.
Specifically, we present numerical results
for a finite chain of DSs supported by an external harmonic trap giving
rise to dynamics that are akin to the oscillations of the
classical Newton's cradle.
Finally, in Sec.~\ref{sec:conclu} we summarize
our results and present possible avenues for further research.

\section{Dynamical reduction for soliton chains}
\label{sec:reduc}

In this section we summarize the dynamical reduction for chains
of BSs and DSs. Under suitable conditions, {\em both} systems
reduce to a Toda lattice on the position of the solitons.
Hence, by initializing the soliton train's positions and
velocities according to the Toda lattice soliton, a travelling
compression pulse can be sustained.

\subsection{Bright soliton chains}

The BS solution (\ref{eq:BS}) describes the coherent evolution of
a density heap on a zero background in an attractive quasi-1D BEC
in the absence of an external confining potential.
When an external trapping potential is included and/or in the
presence of other BSs, the BS is perturbed inducing a deformation
of its shape.
However, under small perturbations,
and noting that BSs are robust, it is possible to approximately
describe the dynamics of the BS by the ansatz (\ref{eq:BS})
as long as its
amplitude, width, position, velocity and phase are dynamically
adjusted to follow accurately the actual solution of the system.
For instance, in the presence of a magnetic trap of the form
\begin{equation}
\label{eq:VMT}
V_{\rm MT}(x) =\frac{1}{2}\, \Omega^2\,x^2,
\end{equation}
where the strength of the trap $\Omega$ is small (compared to the
soliton width), a single BS
solution will undergo left-to-right periodic oscillations of frequency
$\Omega$~\cite{am,sb,st1,st2}.
On the other hand, the presence of another BS, provided that both solitons have
similar amplitudes and velocities and that their separation is large compared
to their widths [ensuring that their shape can still be approximated by
Eq.~(\ref{eq:BS})], their interaction dynamics can be reduced to a set
of coupled ordinary differential equations
(ODEs)~\cite{Karpman:81a,Karpman:81b,Gerdjikov:96,Gerdjikov:97,Arnold:98,Arnold:99}. These reduced ODEs depend on all the parameters of the solitons.
Namely, defining
a vector of time dependent parameters $P_i=(a_i,\xi_i,v_i,\phi_i)$ for
the $i$-th soliton containing, respectively, its amplitude, position,
velocity and phase, the dynamics for the BSs can be approximately
described (under the above mentioned conditions) by a set of
coupled ODEs on the parameters $P_i$ as follows
\cite{Karpman:81a,Karpman:81b,Gerdjikov:96,Gerdjikov:97}:
\begin{equation}
\label{4N-KS}
\left\{
\begin{array}{rcl}
\dot a_j   &=&  4 a_j^2 (S_{j,j-1}-S_{j,j+1}),\\[2.0ex]
\dot v_j   &=& -4 a_j^2 (C_{j,j-1}-C_{j,j+1}),\\[2.0ex]
\dot \xi_j &=&    v_j- 2(S_{j,j-1}+S_{j,j+1}),\\[2.0ex]
\dot \delta_j&=& \frac{a_j^2+v_j^2}{2} - 2 v_j (S_{j,j-1}+S_{j,j+1})\\[1.0ex]
                 & & \,+\, 6\, \nu_j (C_{j,j-1}+C_{j,j+1}),
\end{array}
\right.
\end{equation}
where
\begin{equation}
\label{S-C}
\left\{
\begin{array}{rcl}
S_{j,n}   &=& e^{-|a_n(\xi_j - \xi_n)|} a_n \sin(s_{j,n} \phi_{j,n}),\\[2.0ex]
C_{j,n}   &=& e^{-|a_n(\xi_j - \xi_n)|} a_n \cos(        \phi_{j,n}),\\[2.0ex]
\phi_{j,n} &=& \delta_j - \delta_n - v_n(\xi_j - \xi_n),\\[2.0ex]
s_{j,j-1} &=& 1 = - s_{j,j+1}.
\end{array}
\right.
\end{equation}
%
%
For our particular consideration, we are interested in a chain of
well-separated, almost identical ($a_i \approx a$), BSs.
Under these conditions, the equations of motion for
an infinite chain of ordered ($\xi_{i}<\xi_{i+1}$) BSs can be further simplified
to (see Ref.~\cite{bec-pra} and references therein)
%
\begin{equation}
\ddot\xi_i =  \sigma_{i-1,i}\, 4a^3\,e^{-a(\xi_{i}  -\xi_{i-1})}
            - \sigma_{i,i+1}\, 4a^3\,e^{-a(\xi_{i+1}-\xi_{i})},
\label{eq:BS_ODE1}
\end{equation}
where $\sigma_{i,j}=\pm 1$ is determined by the relative phase of consecutive
($|i-j|=1$) BSs. Namely, $\sigma_{i,j}>0$ corresponds to out-of-phase (OOP)
BSs ($|\phi_j-\phi_i|=\pi$) and $\sigma_{i,j}<0$ corresponds to in-phase (IP)
BSs ($\phi_j-\phi_i=0$). This means that OOP BSs experience mutual repulsion
while IP BSs experience mutual attraction. It is evident that
a homogeneous, equidistant, chain of BSs described by Eq.~(\ref{eq:BS_ODE1})
is a fixed point of the system.
Furthermore, it is straightforward to prove that this fixed point in
the reduced model is unstable in the case of IP BSs and it
is neutrally stable for OOP BSs.
However, stability of the fixed point in the reduced model does
not imply stability of the steady state of the original GP system.
In fact, stability of the reduced model is a necessary but not sufficient
condition for stability of the original GP system.

\begin{figure}[t] 
\begin{center}
\hskip-0.0cm
\psfig{figure=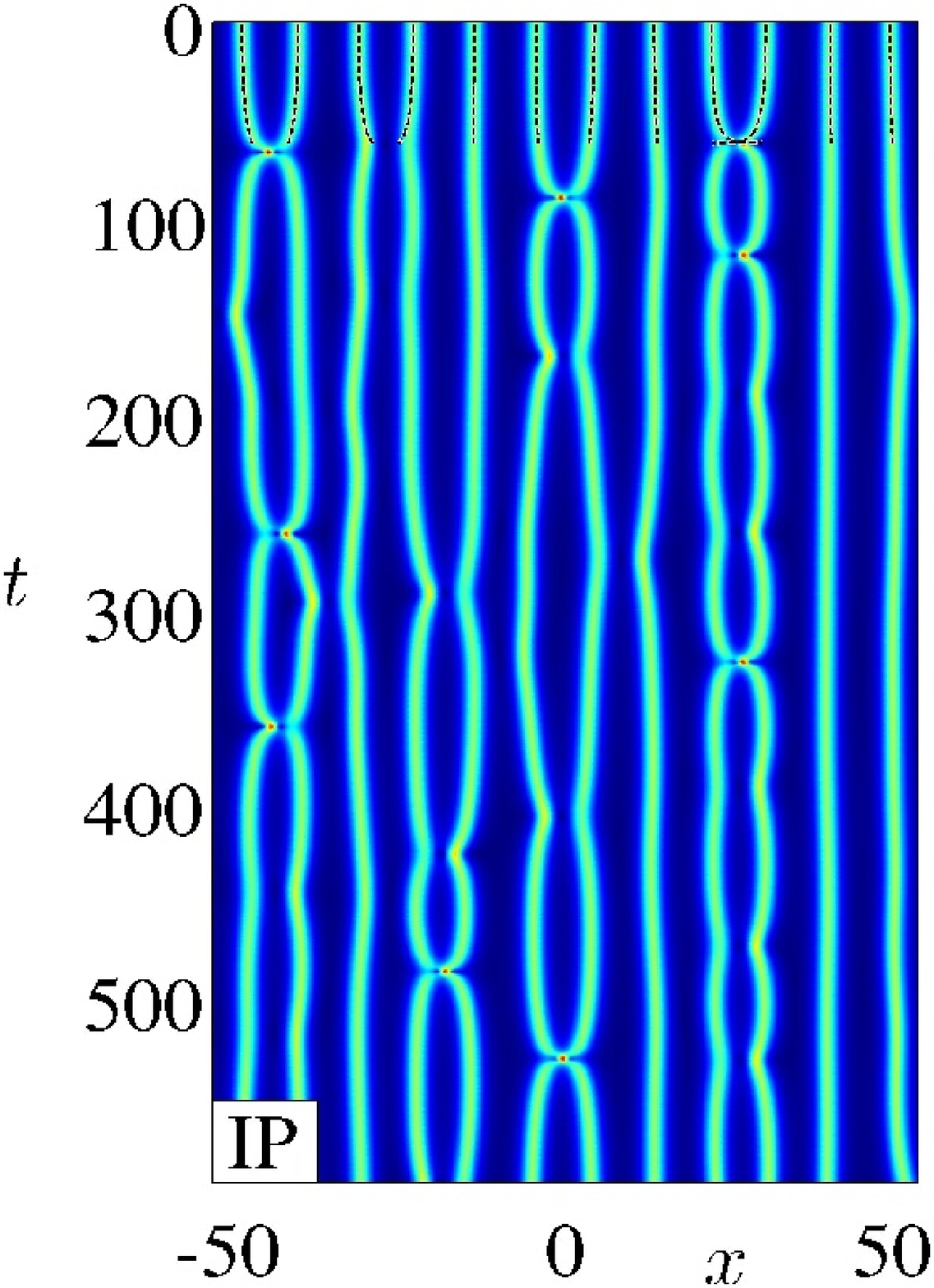, height=5.2cm,silent=}
\psfig{figure=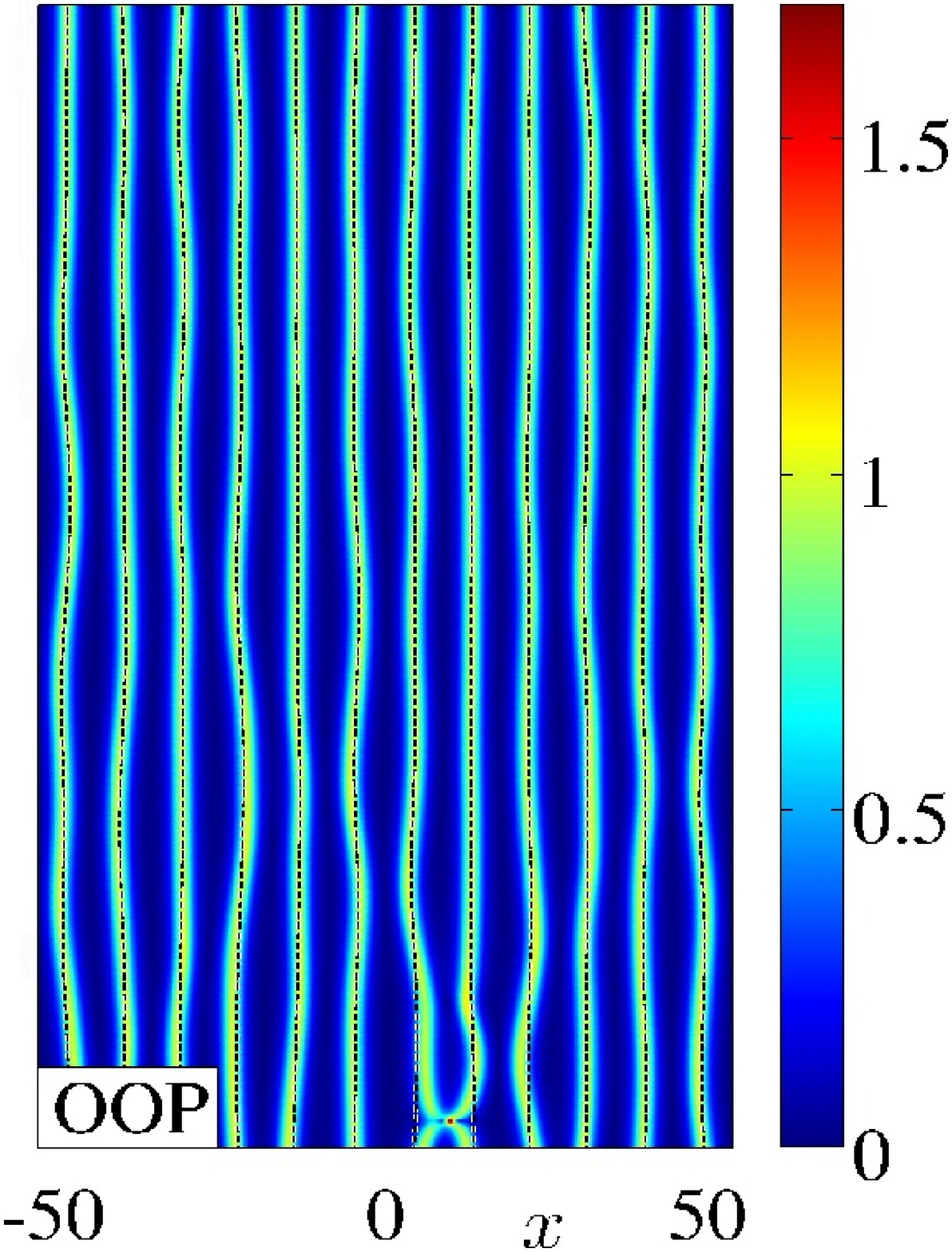,height=5.2cm,silent=}
\\
\psfig{figure=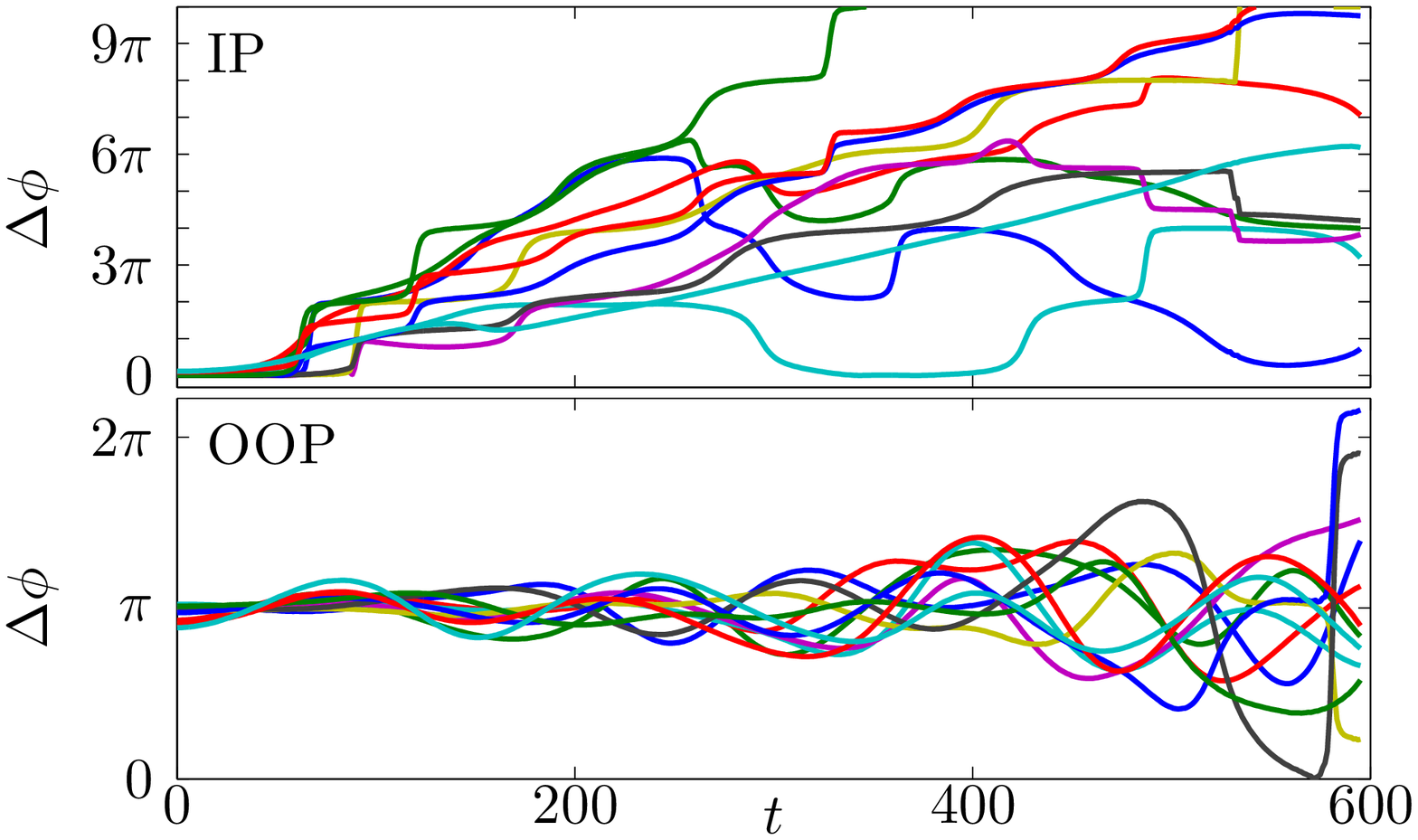,width=8cm,silent=}
\end{center}
\caption{{Color online}.
Dynamics of a slightly perturbed equidistant chain of 12 BSs of
amplitude $a_i=1$ in the periodic domain $x\in[-54,54]$, namely
a separation of $r_0=9$.
The top left and right panels depict, respectively, the spatio-temporal
evolution of the (square root of the) density $|u(x,t)|$ for the
IP and OOP initial chains. The initial conditions are set as the numerically
exact steady states found by Newton iterations and then the BSs positions are
perturbed with randomly distributed displacements between $-r_0/18$ and
$r_0/18$.
The dashed lines represents the orbits obtained from the
corresponding reduced Toda lattice model (\ref{eq:BS_ODE2})
(for the IP case this model can only be integrated until the first
collision time for $t\lesssim 50$).
The bottom two panels depict the phase difference between consecutive
density maxima. These correspond to the relative phases $\Delta\phi$
between consecutive BSs (top and bottom subpanels corresponding,
respectively, to the IP and OOP cases).
}
\label{fig:u2_BS_random}
\end{figure}

Since we are interested in excitations of the homogeneous chain, let us first
study their stability. In Fig.~\ref{fig:u2_BS_random}, we depict typical time
evolutions corresponding to an IP (top-left) and OOP (top-right) BS chain.
The system is initialized using the numerical steady state of the
GP system with periodic boundary conditions ---found using a standard fixed
point iteration algorithm--- with a small perturbation.
As can be seen in the figure, both the IP and OOP cases are unstable.
The nature of the instability seems, however, different. The IP case, for which
we know that even in the reduced ODE model is unstable due to the
mutual attraction between BSs, seems to be strongly unstable.
The instability is
manifested by two neighboring BS coalescing as early as $t\approx 50$.
This instability is easy to understand since a small perturbation will induce
two neighboring BSs to be slightly closer than its other neighbors,
thus accelerating the process of attraction and hence leading to a
rapid collision between these two BSs.
On the other hand, in the OOP case, the reduced chain is neutrally stable
and, thus, perturbations with respect to the positions of the BSs from the
equidistant chain should not cause instabilities.
As shown in the top-right panel of Fig.~\ref{fig:u2_BS_random}, this is the
case for intermediate times ($t<500$) where the mutual repulsion between BSs
is responsible for collision avoidance between neighboring BSs.
Nonetheless, as the panel shows, for later times, $t \approx 575$, a collision
between neighboring BSs does indeed occur.
The presence of a collision is unequivocal evidence that the involved BSs
where not OOP when they collided.
In fact, the loss of the OOP property between consecutive
BSs is precisely what induces the instability of the otherwise OOP initial
BS chain.
The desynchronization between the phases can be clearly seen in the bottom two
subpanels of Fig.~\ref{fig:u2_BS_random}. The panels depict the time
evolution of the relative phase between consecutive BSs that initially start
in the IP (top subpanel) and the OOP (bottom subpanel) configurations.
It is clear that the OOP property between consecutive BSs
is approximately held for intermediate
times (see $t<500$ in the bottom subpanel) ensuring the mutual repulsion
between consecutive BSs and, thus, stability for the chain. However, it is also
clear that the OOP property gets progressively worse until a pair of
consecutive BSs have a {\em zero} relative phase between them, i.e.~they
are IP around $t=575$, inexorably leading to their collision.

\begin{figure}[t] 
\begin{center}
\hskip-0.5cm
\psfig{figure=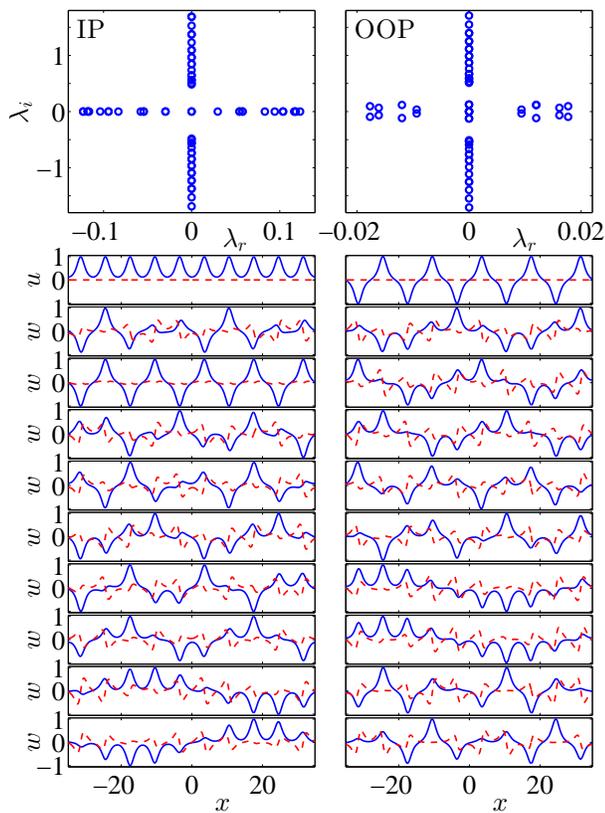,width=8cm,silent=}
\end{center}
\caption{{Color online}.
BdG stability spectrum for the steady state for a chain of 10 equidistant chain of
IP (top left) and OOP (top right) BSs in the periodic domain $x\in[-35,35]$.
The bottom set of panels depict, from top to bottom, the steady state and
the first nine most unstable eigenfunctions where the real part is depicted
by the (blue) solid line and the imaginary part by the (red) dashed line.
}
\label{fig:stab_BS}
\end{figure}

To further understand the nature of the instability, we have computed the
Bogolyubov-de Gennes (BdG) stability spectrum for the IP and OOP steady states
of the GP system. The BdG spectrum is computed by considering perturbations
from the steady state $u_0(x)$ of the following form:
\begin{equation}
u(x,t) = \left( u_0(x) + \varepsilon (a(x) e^{\lambda t} + b(x) e^{-\lambda^* t}) \right)
e^{-i\mu t},
\label{eq:BdG}
\end{equation}
where $\mu$ is the so-called chemical potential (or [the negative of] the
temporal frequency) of the solution, $\lambda$ is the eigenvalue with
associated eigenvector $\{a,b^*\}$, and where $(\cdot)^*$ stands for
complex conjugation. After applying the perturbation ansatz (\ref{eq:BdG})
into the GP equation (\ref{GPE}) and linearizing the ensuing equation,
one obtains an eigenvalue problem with a corresponding eigenfunction
$w(x)=a(x) + b^*(x)$ at $t=0$.
After computing the spectrum, any eigenvalue $\lambda=\lambda_r + i\lambda_i$
with a positive real part ($\lambda_r$) indicates an unstable eigenfunction.
The spectrum associated with the IP and OOP steady states is depicted,
respectively, in the left and right top panels of Fig.~\ref{fig:stab_BS}.
As is can be noticed, the OOP spectrum has a handful of complex eigenvalues
with a {\em small} ($\lambda_r<0.02$) real part indicating a weak instability.
In contrast, the IP case reveals a larger (purely real) instability with
$\max(\lambda_r)\approx 0.123$, indicating a stronger instability.
Closer inspection of the unstable eigenfunctions (see bottom set of panels
in Fig.~\ref{fig:stab_BS}) reveals that the instabilities manifest themselves as
local translational modes for consecutive BSs in the opposite direction and,
thus, bringing them closer to each other.
We have checked that the stability results above
(cf.~Figs.~\ref{fig:u2_BS_random} and~\ref{fig:stab_BS}) are very similar
for other values of the parameters such as amplitude, number, and separation of the BSs,
as well as different domain lengths. Evidently, as more BSs are included in the
system, a higher degeneracy of the eigenvalues arises since all solutions and
eigenfunctions posses translational symmetry.
Finally, it is relevant to mention that the BS chain might be rendered stable
by a suitable choice of periodic lattice potential providing stabilizing
pinning for each BS located at the respective minimum of the lattice
potential~\cite{pre-foc,bec-pra}.
However, we do not explore this avenue further in this manuscript.

Since the IP BS chain is highly unstable, we will focus our attention in
the case of the weakly unstable OOP chain.
Therefore, let us consider the case $\sigma_{i,j}=+1$ for which the
reduced equation of motion yields
\begin{equation}
\ddot\xi_i =  4a^3\,e^{-a(\xi_{i}  -\xi_{i-1})}
            - 4a^3\,e^{-a(\xi_{i+1}-\xi_{i})},
\label{eq:BS_ODE2}
\end{equation}
which has precisely the form of the celebrated Toda lattice~\cite{Toda_book}
that is further described below.
It is important to mention at this stage that the restriction of locked phases
between BSs (IP or OOP) is not necessary to obtain a Toda lattice-type model.
In fact, by allowing the phase of each BS to dynamically evolve, the equations
of motion (\ref{4N-KS}) reduce to a {\em complex} Toda lattice of
the form \cite{Gerdjikov:97}:
\begin{equation}
\label{CTL}
{\ddot q_j} = 2a\, \left(e^{-(q_j - q_{j-1})} - e^{-(q_{j+1} - q_j)} \right),
\end{equation}
where
the corresponding complex
variable $q_j$ for this complex Toda lattice is defined through
the original BS's parameters by
\begin{equation}
\label{qj}
q_j =  a\, \xi_j - j\,\ln{(a^2)} + i\, \left(j\,\pi - v\xi_j + \delta_{j} + \delta \right),
\end{equation}
where $a$ and $v$ are the ensemble average height and velocity of
the BSs, while $\xi_j$ is the position of the $j$-th BS and
the $\delta_j$'s are the BS phases and $\delta$ is their average.

\subsection{Dark soliton chains}

As was done in the previous section for the BS chain,
the case of DS chains can also be reduced to
a set of ODEs on the DS parameters. This reduction is obtained through
DS perturbation theory as described in detail in
Refs.~\cite{Kivshar_BAM_1989,Kivshar_Pang_94,Kivshar_Krol_95,DJF_REVIEW,DARK_BOOK}.
In particular, considering a DS ansatz of the
form (\ref{eq:DS}) for all DSs in a chain supported on
a constant background with density $n_0$, the equations of motion are
approximately reduced to
\begin{equation}
\ddot\xi_i = 8n_0^{3/2}\,e^{-2\sqrt{n_0}(\xi_{i}  -\xi_{i-1})}
           - 8n_0^{3/2}\,e^{-2\sqrt{n_0}(\xi_{i+1}-\xi_{i})},
\label{eq:DS_ODE}
\end{equation}
which, as for the BS chain, is a form of the Toda lattice~\cite{Toda_book}
on the DS positions.

The main difference between the reduced dynamics of
bright and dark solitons that is
worth pointing at this stage is that BSs can have mutual interactions
that are repulsive or attractive depending on their relative phases
as described above. On the other hand, DSs are {\em always} repulsive.
Therefore, the stationary homogeneous, equidistant, DS chain is {\em always}
stable. This observation will be crucial when comparing the dynamics
of BS and DS chains from the GP model (\ref{GPE}) and their respective
dynamical reductions (\ref{eq:BS_ODE1}) and (\ref{eq:DS_ODE}).

\subsection{Validation of the Toda lattice reduction}
%
We now validate the dynamics obtained from the dynamical reduction
for both bright and dark soliton chains. In order to numerically approximate
an infinite lattice, we take a periodic domain in the interval $x\in[-L,L]$ and
place the solitons equidistant from each other accounting for the periodic
boundary conditions.
Thus, considering $N$ solitons gives a steady state equidistant
configuration such that $|\xi_{i+1}-\xi_i|=r_0$ for $i=1,\dots,N-1$,
where the distance $|\cdot|$ is measured in the periodic domain
such that $|\xi_{N}-\xi_1|=2L-r_0$.

Let us first consider a BS chain.
%
As we described earlier, the perturbed equidistant chain evolves as depicted in
Fig.~\ref{fig:u2_BS_random} where the top left and
right panels correspond, respectively, to the IP and OOP cases.
We have superimposed the corresponding dynamics of the reduced ODE model
of the BSs (\ref{eq:BS_ODE2}) using dashed lines.
As can be observed from the figure, in the IP case (top left panel)
before the collision of the BSs ($t<50$), the ODE model
very closely follows the BS dynamics. After collision, the
ODE model breaks down as the BS centers coalesce and, thus,
we only show the reduced ODE orbit up to the first collision time.
On the other hand, for intermediate times, the OOP chain (top right panel)
does not suffer from the collision of BSs as it assumes that all BSs
are {\em always} OOP. The resulting reduced ODE dynamics closely
follows the original GP dynamics for short times, but later
deteriorates since, as explained earlier, the BSs lose the OOP
synchronization.
Nonetheless, for intermediate times, while the BSs are kept
separated, the reduced ODE model does reproduce the original GP dynamics.

\begin{figure}[t] 
\begin{center}
\psfig{figure=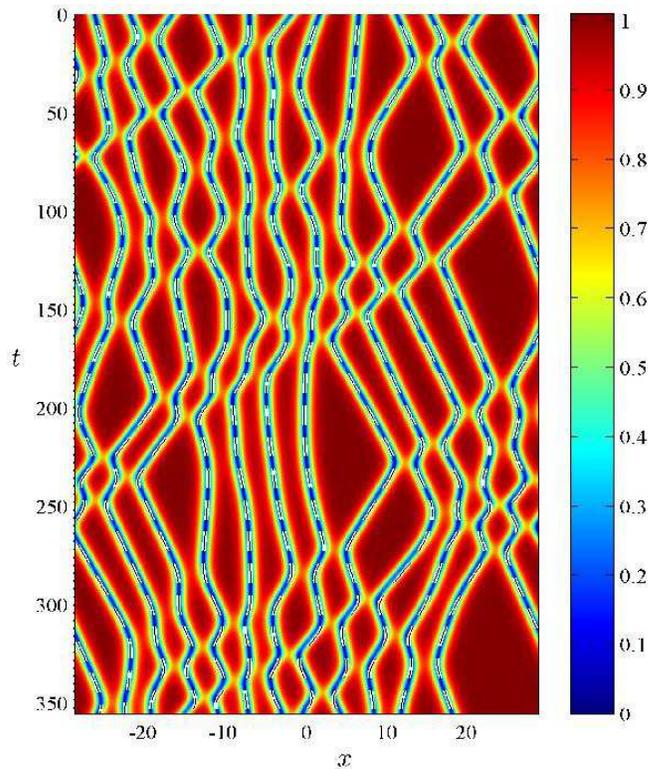,width=8.5cm,silent=}
\end{center}
\caption{{Color online}.
Interaction dynamics for 12 DSs initialized at random positions
with random velocities in a periodic domain. Depicted is the
spatio-temporal evolution of the  (square root of the) density.
The dashed lines represent the orbits obtained from the
corresponding reduced Toda lattice model (\ref{eq:DS_ODE}).}
\label{fig:u2_random}
\end{figure}

In contrast to the BS chain, the DS chain does not suffer from
phase-induced instabilities since DSs {\em always} repel each other.
As a result, the reduced
ODE model for the DS chain (\ref{eq:DS_ODE}) provides a very robust
model for the GP dynamics under extended time evolutions for {\em any}
initial condition provided the DSs are initially well-separated.
An example of the dynamics from the reduced Toda lattice ODE (\ref{eq:DS_ODE}) and
the original GP model is depicted in Fig.~\ref{fig:u2_random}, where a collection
of DSs placed at random locations with random initial velocities evolves
in time. It is clear from the figure that the Toda lattice model gives
an accurate prediction of the DS positions for the
original GP system for long times.

\section{Toda lattice solitons}
\label{sec:TL}

\subsection{Preliminaries}

Before constructing Toda lattice solitons on the chains of bright
and dark solitons, let us review the form of these solutions for
completeness.
The Toda lattice is one of the most popular models in physics since,
by construction, it was designed as to prescribe
a chain of {\em nonlinear} oscillators with completely integrable
evolution~\cite{Toda_book}. As such, the Toda lattice possesses some exact solutions
that are the foundation for building more complex solutions. In particular,
the Toda lattice possesses periodic and localized solutions \cite{Toda_book}.
Here we focus on the latter type of solutions referred to as Toda solitons.
The Toda lattice's equations of motion
\begin{equation}
\begin{array}{rcl}
\ddot y_n &=&  V_{\rm TL}(y_{n+1}-y_n) - V_{\rm TL}(y_{n}-y_{n-1}), \\[2.0ex]
          &=&  A\,e^{-b(y_{n}  -y_{n-1})} - A\,e^{-b(y_{n+1}  -y_{n})},
\end{array}
\label{eq:TL}
\end{equation}
originate from the interaction of nearest neighbors in a
one-dimensional chain of coupled, unit mass, particles at positions $y_n$,
interacting through the potential
\begin{equation}
V_{\rm TL}(\Delta y)  =\frac{A}{b}e^{-b\,\Delta y}+A\,\Delta y.
\label{toda potential}
\end{equation}
Here, $\Delta y$ is the separation between particles and $A$ and $b$ are
positive parameters prescribing, respectively, the strength and decay of the
inter-particle interactions.
By following the evolution of the particles through their mutual separation
\begin{equation}
\Delta y_{n}=y_{n+1}-y_{n},
\label{eq:sep}
\end{equation}
and defining $s_{n}\equiv d({{\Delta y}}_{n})/dt$ and $p_{n}\equiv dy_{n}/dt$
so that $s_{n-1}-s_{n}=p_{n}$, the equations of motion
can be rewritten in term of $S_{n}=\int s_{n}dt$ as
\begin{equation}
\ln\left(  1+\frac{\ddot{S}_{n}}{a}\right)  =\frac{b}{m}\left(  S_{n+1}%
-2S_{n}+S_{n-1}\right).
\label{toda in dual}
\end{equation}
Then, it is straightforward to find solitary kink solutions for this system
in the form:
\begin{equation}
s_{n}=\pm\frac{\beta}{b}\tanh\left(  n\kappa\pm\beta t\right) +{\rm const},
\label{eq:TL_kink}
\end{equation}
where the kink velocity is $c=\beta/\kappa$ and its amplitude
$\beta$ is given by
\begin{equation}
\beta=\displaystyle\sqrt{Ab}\, \sinh\kappa,
\label{eq:TL_beta}
\end{equation}
where the width of the kink $\kappa$ is a free parameter.
It should be noticed that this solution is stable and it
corresponds to a compression wave that travels through the
lattice~\cite{Toda_book}.

\subsection{Toda lattice solitons: hypersolitons}

We now seek to use the soliton solution for the Toda lattice (see previous
section) to construct a Toda lattice soliton on the reduced lattice equations
for the bright and dark soliton chains.
Let us consider an equilibrium configuration consisting of a chain of
$N$ equidistant solitons with separation $r_0=2L/N$ in the periodic
domain $x\in[-L,L]$.
Both OOP\footnote{From now on, since we are focusing on the
OOP BS case, we may omit the term OOP} BS and DS chains are reduced,
respectively, to the Toda lattice chains (\ref{eq:BS_ODE2}) and (\ref{eq:DS_ODE})
where the Toda lattice potential parameters are given in terms of soliton
amplitude $a$ for the BS chain and in terms of the background density
$n_0$ for the DS chain.
It is worth mentioning that the uniform pre-compression experienced
by the periodic chain effectively corresponds to a rescaling on the
strength of the Toda lattice potential $A$. This is evident when rescaling
the soliton positions by a factor $\gamma$, $y_n=\gamma \tilde y_n$, then the
exponential interaction terms become
$A\,e^{-b(y_n-y_{n-1})}=
       A\,e^{-b\gamma(\tilde y_n-\tilde y_{n-1})}=
\tilde A\,e^{-b      (\tilde y_n-\tilde y_{n-1})}$
where $\tilde A = A\,e^{-b\gamma}$.

\begin{figure}[t] 
\begin{center}
\hskip-0.7cm
\psfig{figure=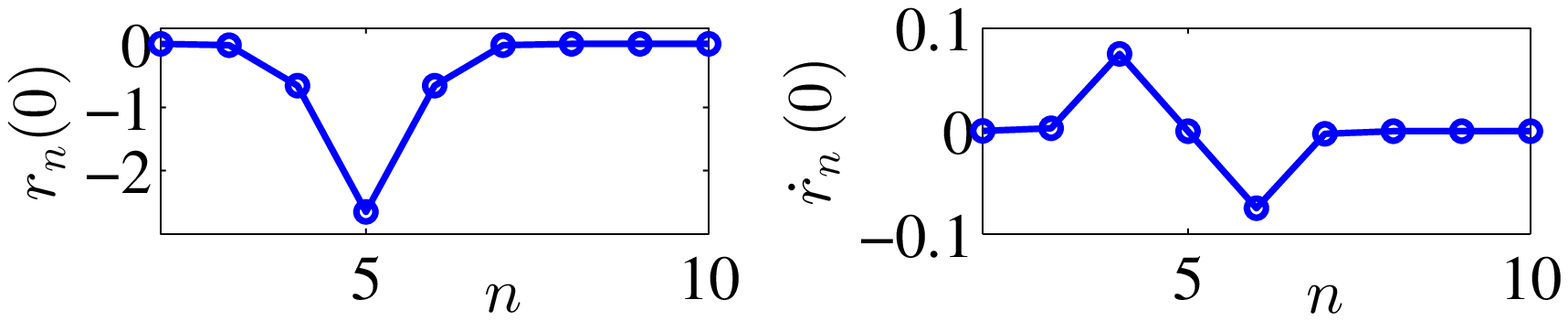,width=7.7cm,silent=}
\psfig{figure=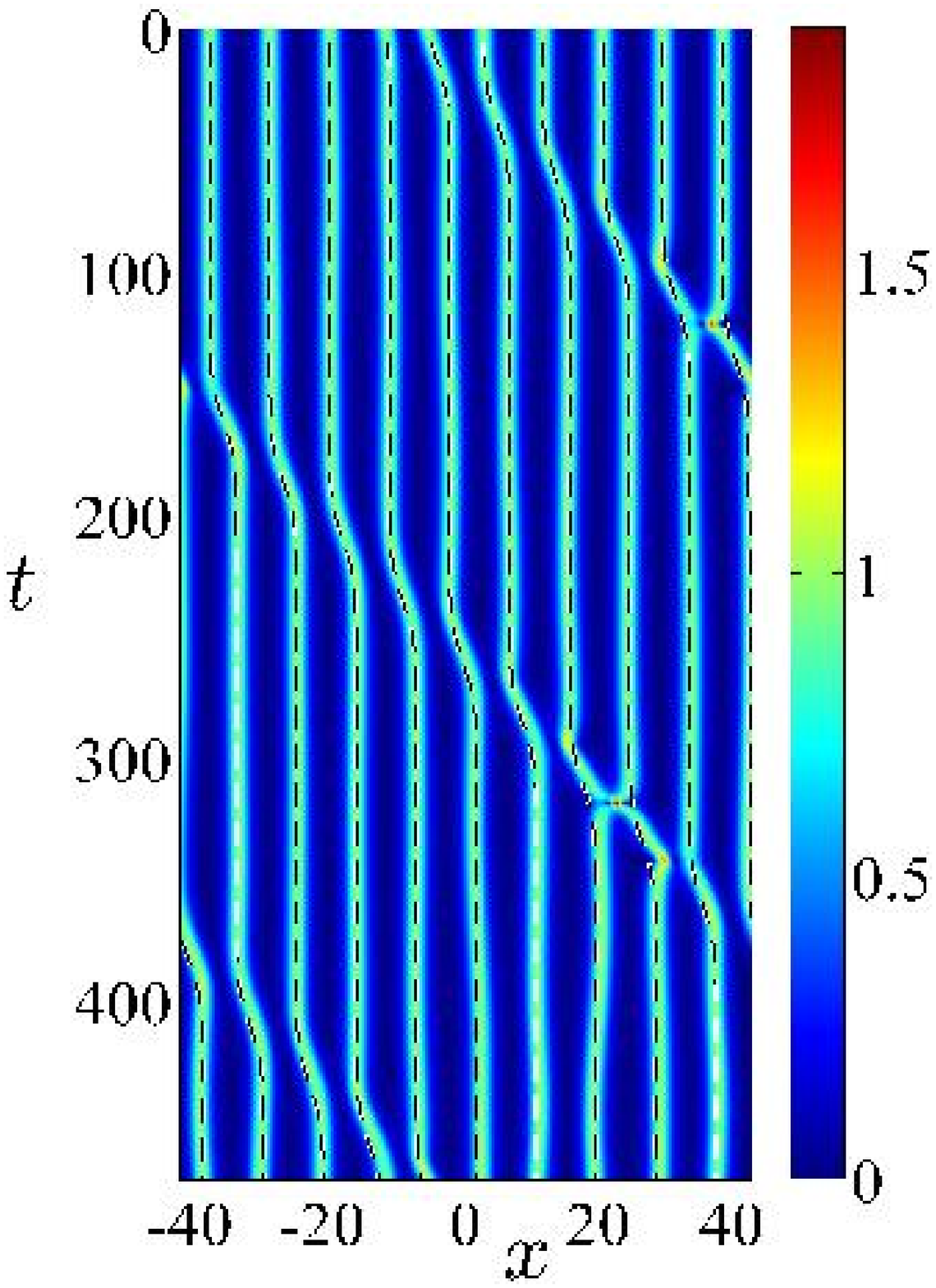, height=5.5cm,silent=}
~
\psfig{figure=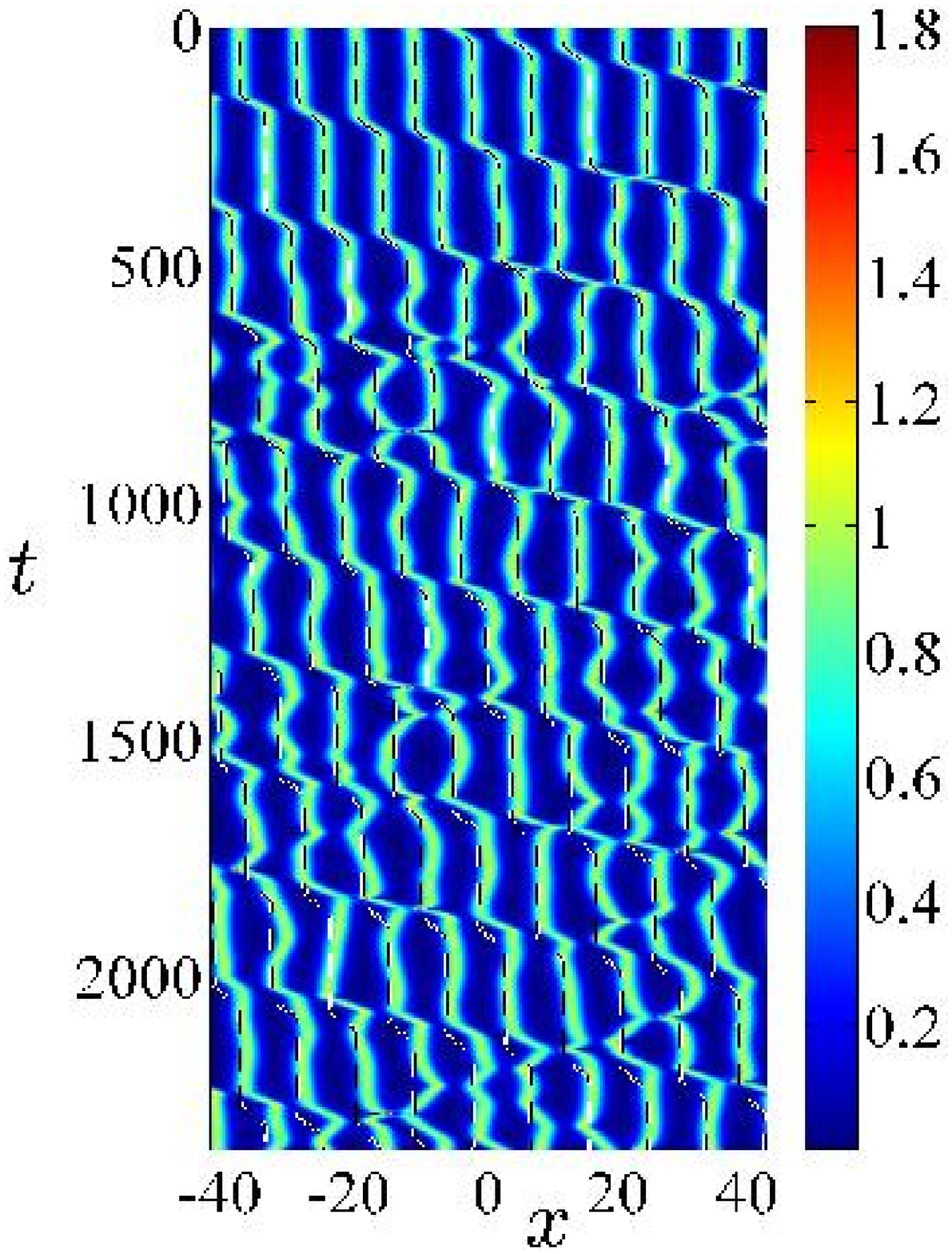, height=5.5cm,silent=}
\end{center}
\caption{{Color online}.
Toda lattice soliton riding on a chain of $N=10$ BSs in the
periodic domain $x\in[-43,43]$. The top panels depict the
initial conditions in terms of the relative displacements from equilibrium
$r_n(t=0)$ (left subpanel) and their corresponding velocities 
$\dot r_n(t=0)$ (right subpanel).
The bottom left and right panels depict the evolution of the 
(square root of the) density
of the original GP model initialized with the Toda lattice soliton
for, respectively, a total time equivalent to two and ten cycles of the
Toda lattice soliton around the periodic domain.
The dashed lines correspond to the reduced ODE model of the Toda lattice
(\ref{eq:BS_ODE2}).
}
\label{fig:u2_BS_TL}
\end{figure}

Let us start by initializing the chain of BSs
such that the initial positions and initial velocities satisfy the
corresponding Toda soliton (\ref{eq:TL_kink}). An example of this case is
depicted in Fig.~\ref{fig:u2_BS_TL} for $N=10$ BSs in the periodic chain
$x\in [-43,43]$. The top panels depict the initial condition
for the displacements from equilibrium between 
solitons $r_n=\xi_n-\xi_{n-1}-r_0$ (left subpanel) and their respective
initial velocities ${\dot r}_n$. As it is evident from these panels, the
kink (\ref{eq:TL_kink}) corresponds to a localized {\em compression}
wave for the soliton's positions.
It is worth mentioning at this stage that contrary to the homogeneous, equidistant,
chain where $2L=N\, r_0$, for the chain initialized with the Toda lattice solitons
the length of the domain has to be adjusted since we are introducing a compression
wave to the initial condition. Thus, we compute the length of the domain by
adding all the separations between consecutive solitons.
The bottom row of panels in Fig.~\ref{fig:u2_BS_TL} depicts the evolution of
the density after seeding the original GP equation with the Toda lattice
soliton initial conditions on a pre-compressed BS chain. As it can be observed
from the left subpanel, the initialized compression wave travels at the
prescribed speed ---for this panel we set the final time to precisely
$2c/(2L)$, namely, the time needed to perform exactly two complete cycles through
the lattice. In the figure, the dashed lines correspond to the Toda lattice
soliton from the reduced ODE model (\ref{eq:BS_ODE2}) which, for short times, accurately
approximates the full GP dynamics.
However, as we have described before, the BSs will eventually lose
their OOP synchronization leading to the coalescence of two consecutive
solitons. The first coalescence occurs approximately at $t=125$ for the 
right-most two
solitons of the lattice. Nonetheless, after this coalescence, the Toda
lattice soliton seems to reform again which, in turn, suffers from the
coalescence of more BSs as time progresses. It is interesting to note that
although the reduced ODE should fail after a BS pair loses its OOP
synchronization, and as it was noted in Ref.~\cite{Gerdjikov:97}, for OOP
initial conditions, the reduced ODE model still captures the dynamics of
the interacting chain for some time. Nonetheless, as can be seen in the
bottom right panel of Fig.~\ref{fig:u2_BS_TL}, after longer integration times
(10 full cycles around the lattice),
the Toda soliton does not preserve its shape and other excitations start
populating the dynamics, including Toda solitons that apparently
move in the opposite direction of the original Toda soliton.
By the same token, it is also evident that the ODE description (see dashed
line) fails to capture the long term dynamics of the original GP system.
This is evidence that the unstable character of the BS chain due to the
desynchronization of the phases precludes satisfactory modeling of the
full GP system with the reduced ODE (\ref{eq:BS_ODE2}) where all BSs
are assumed to be OOP.

\begin{figure}[t] 
\begin{center}
\hskip-0.65cm
\psfig{figure=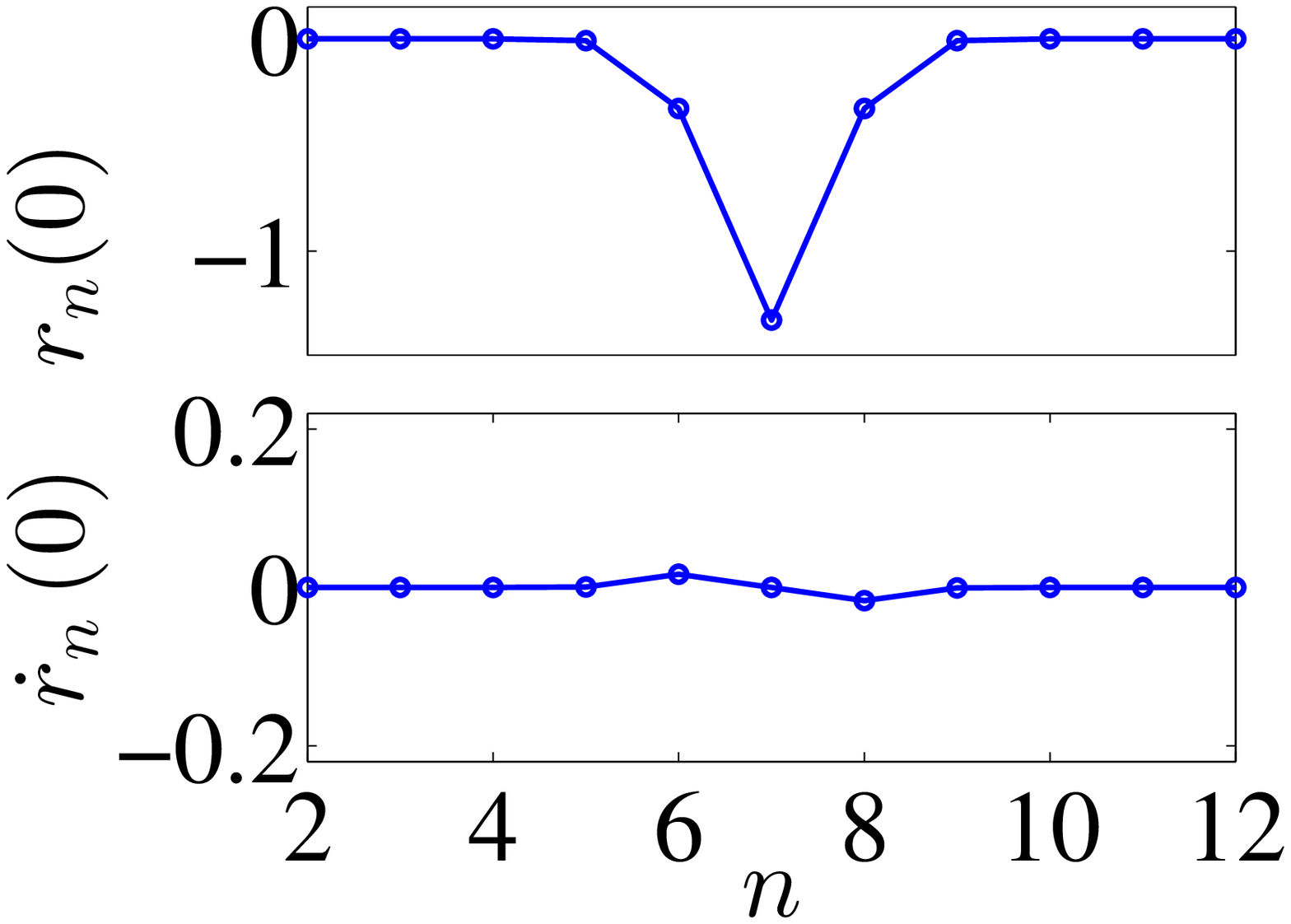,height=2.05cm,silent=} 
\psfig{figure=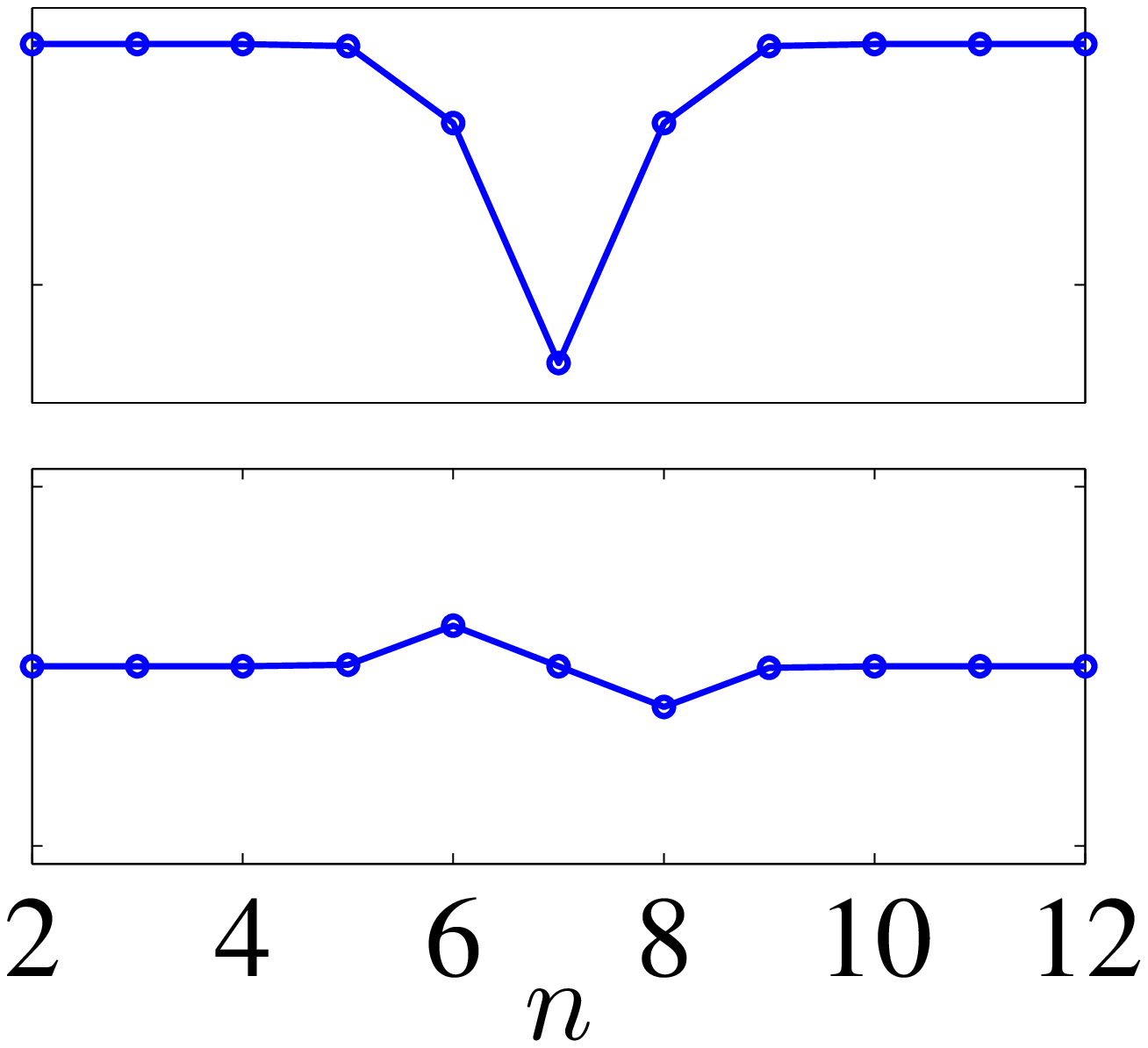,height=2.05cm,silent=} 
\psfig{figure=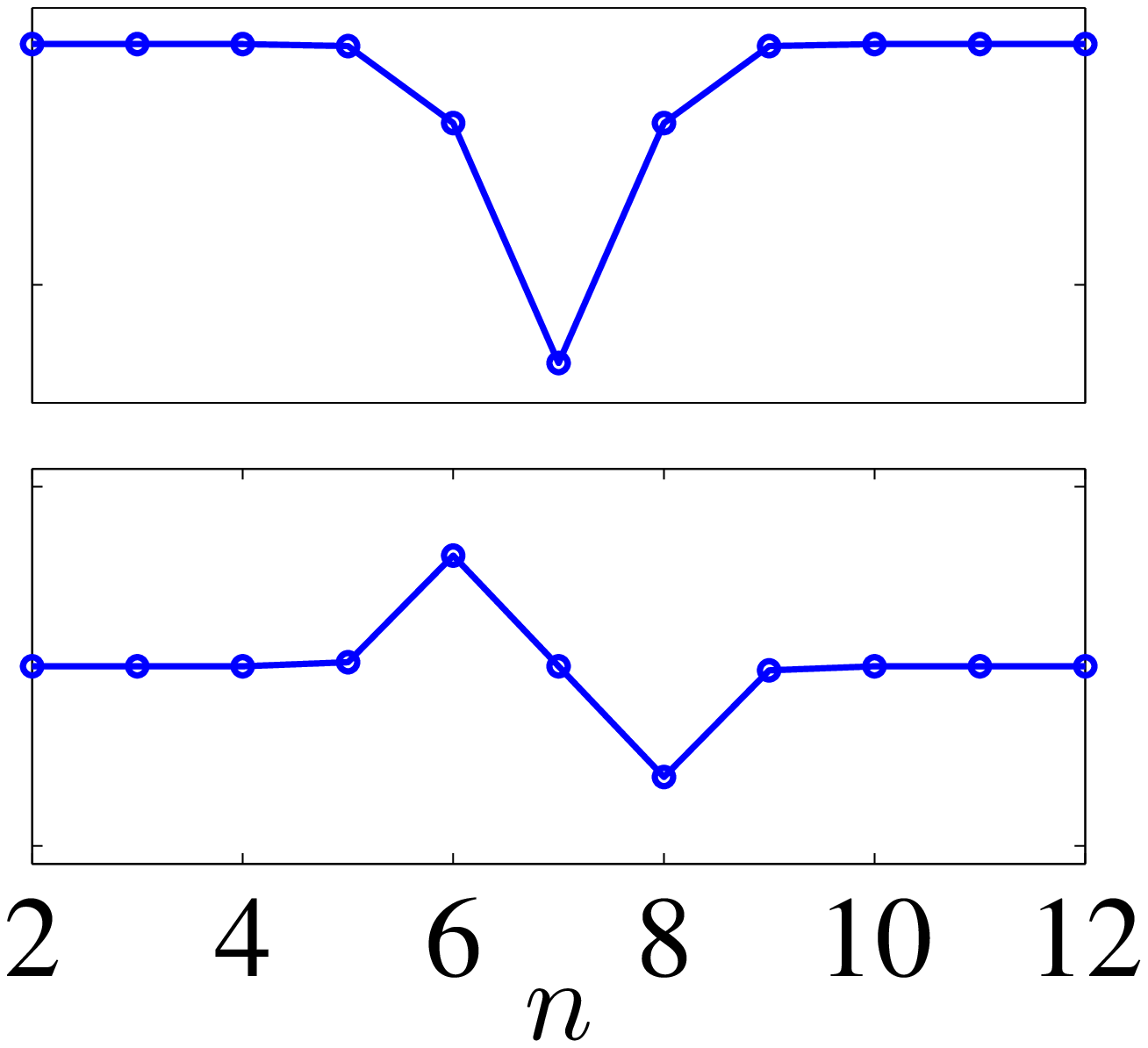,height=2.05cm,silent=} 
\\
\hskip-0.1cm
\psfig{figure=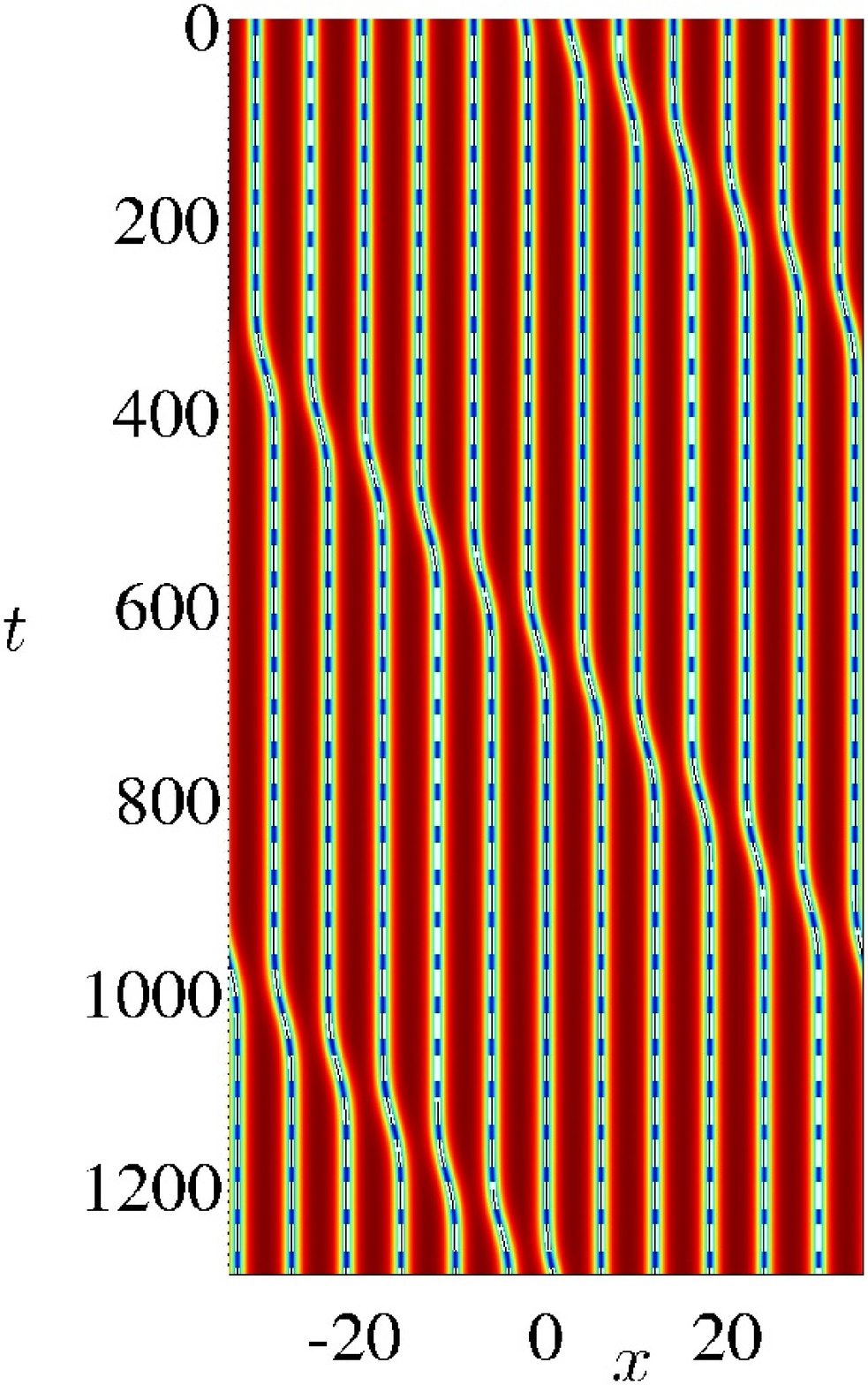,height=4.5cm,silent=} 
\psfig{figure=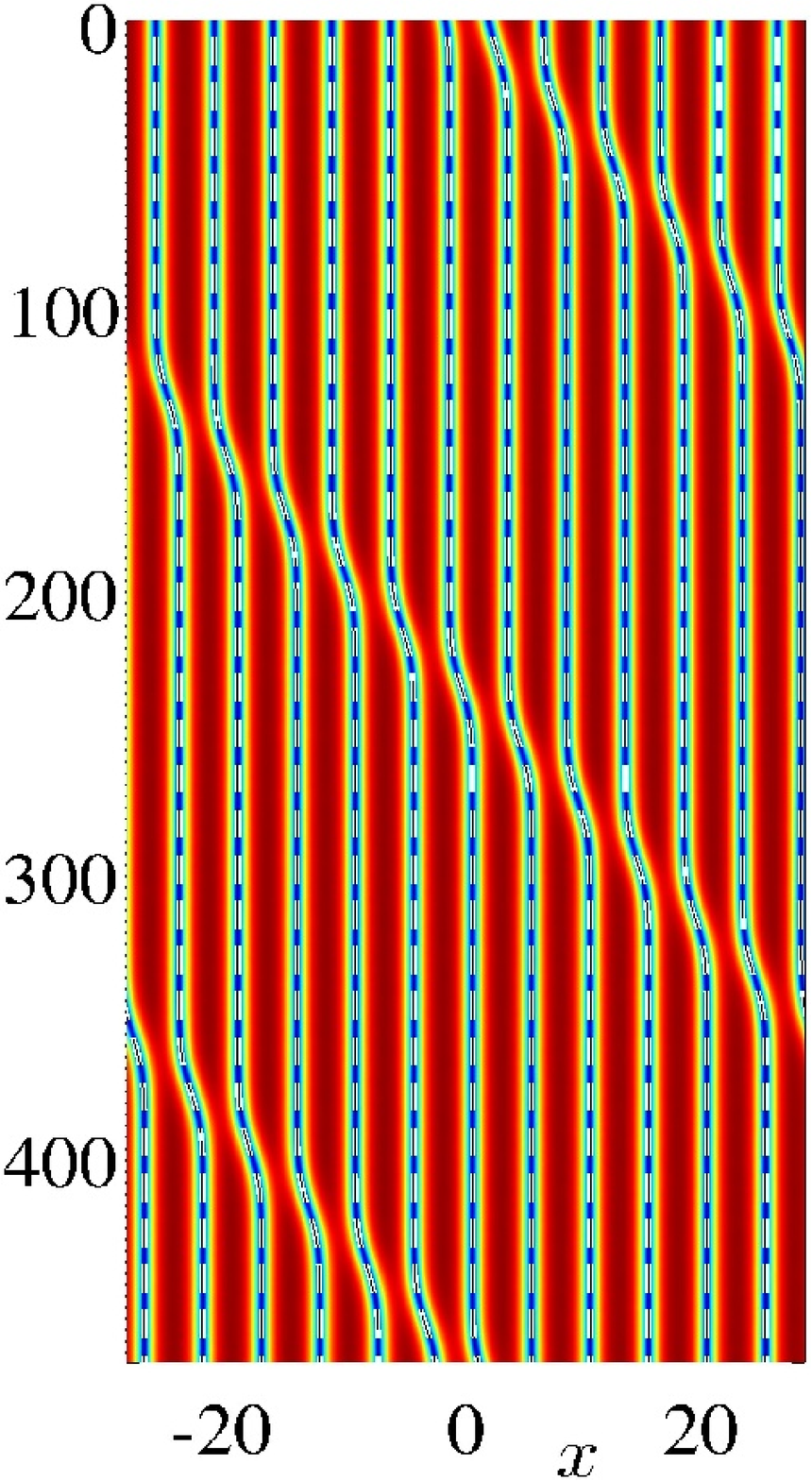,height=4.5cm,silent=} 
\psfig{figure=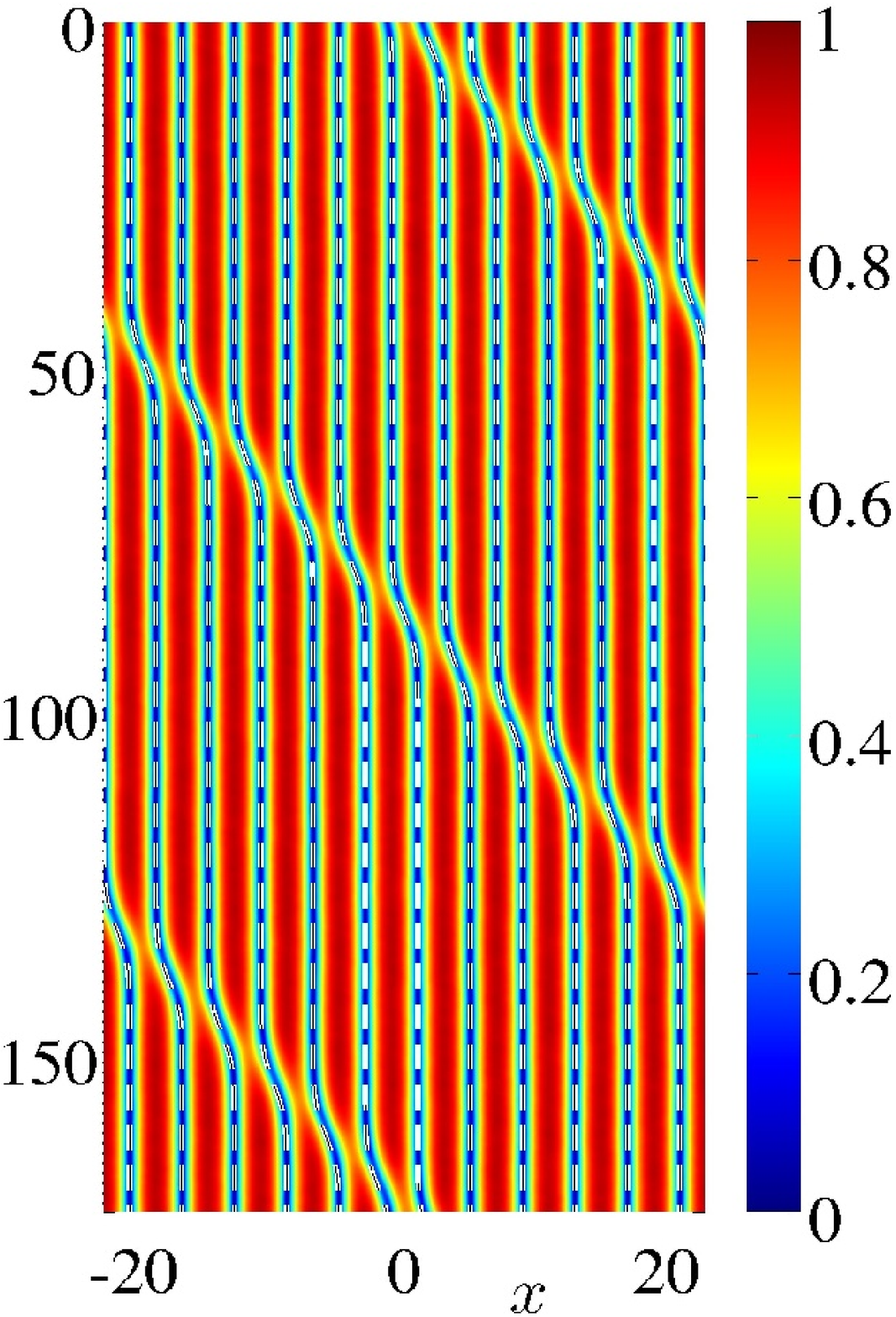,height=4.5cm,silent=} 
\\
\psfig{figure=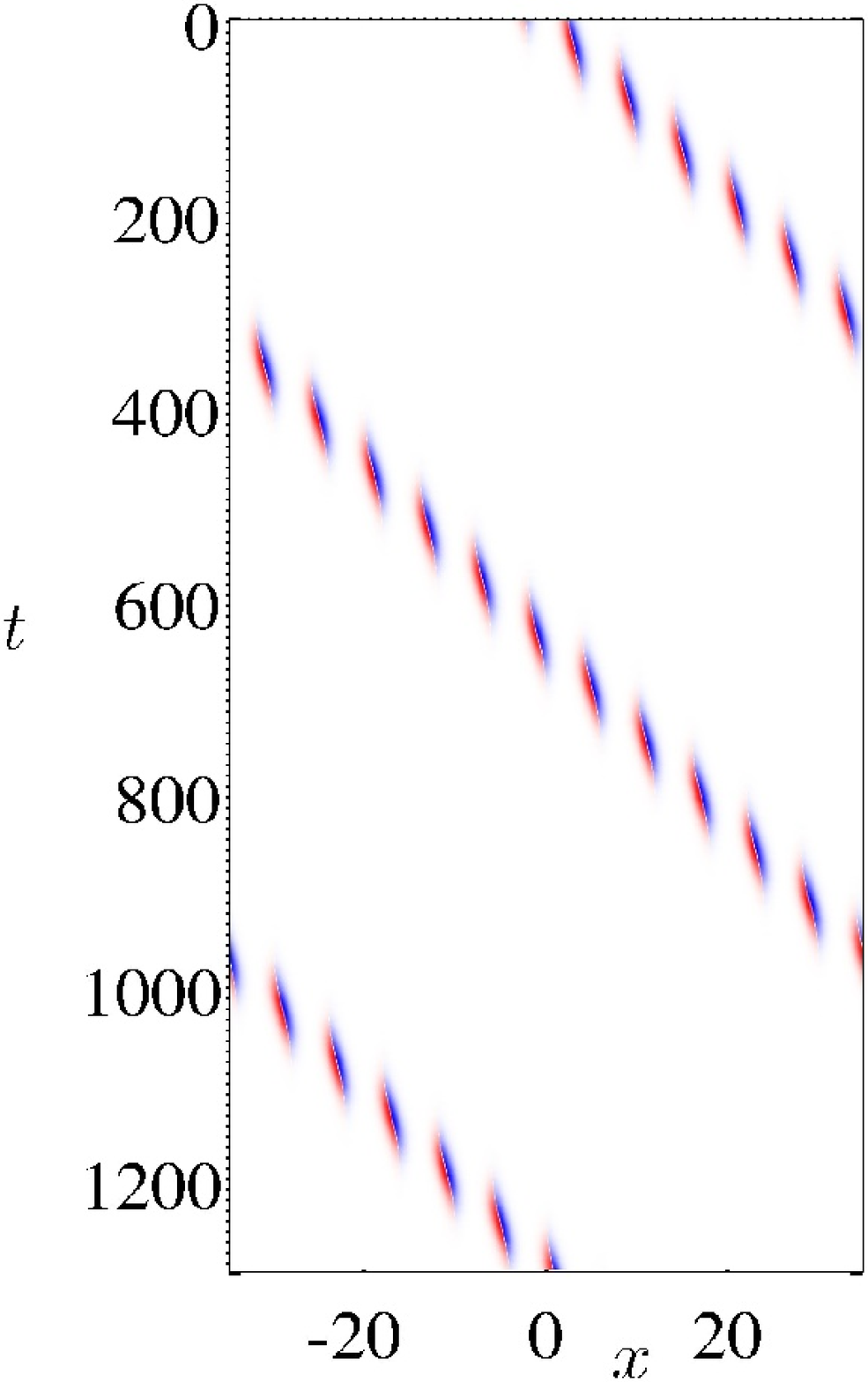,height=4.4cm,silent=} 
\psfig{figure=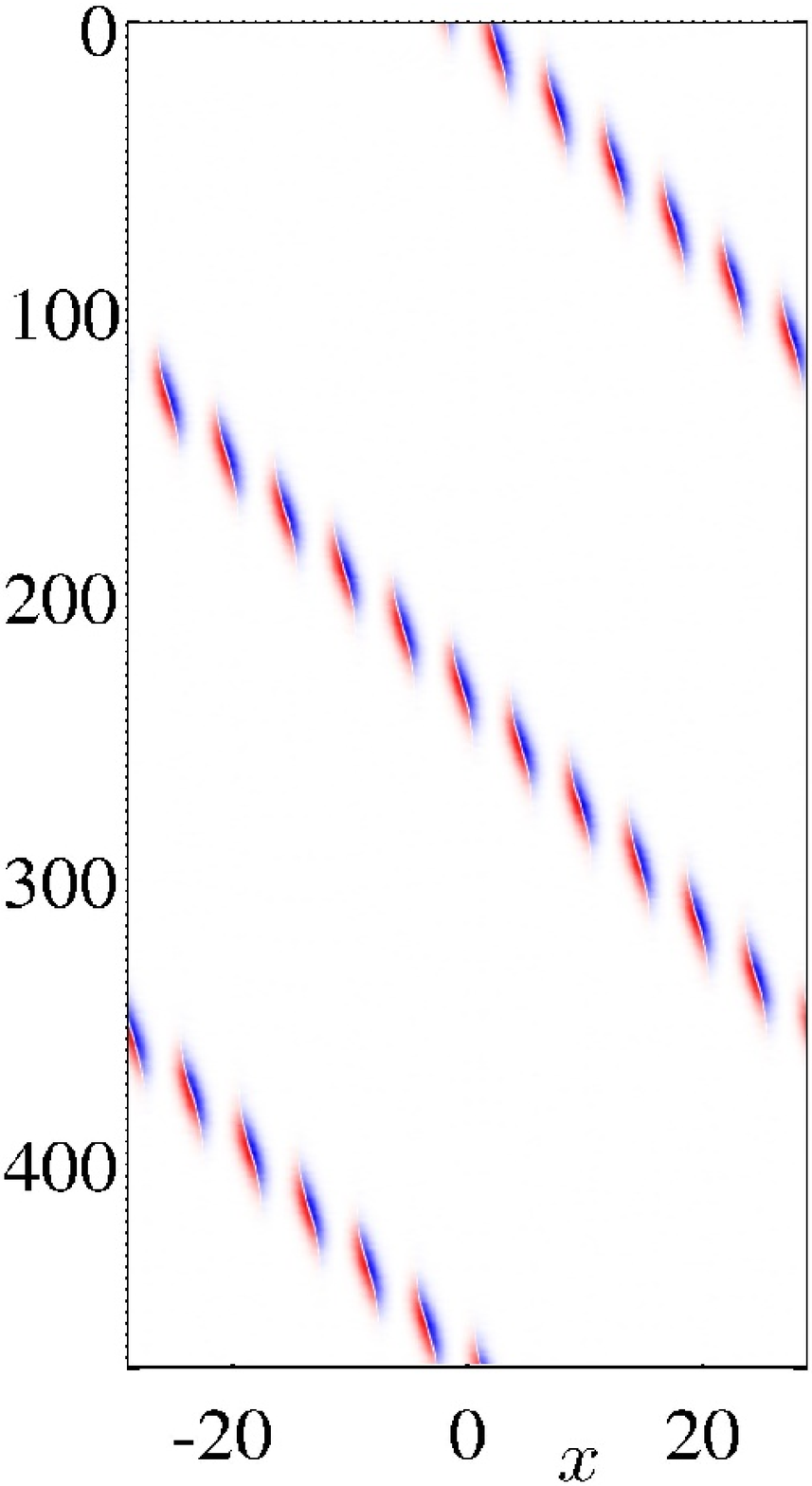,height=4.4cm,silent=} 
\psfig{figure=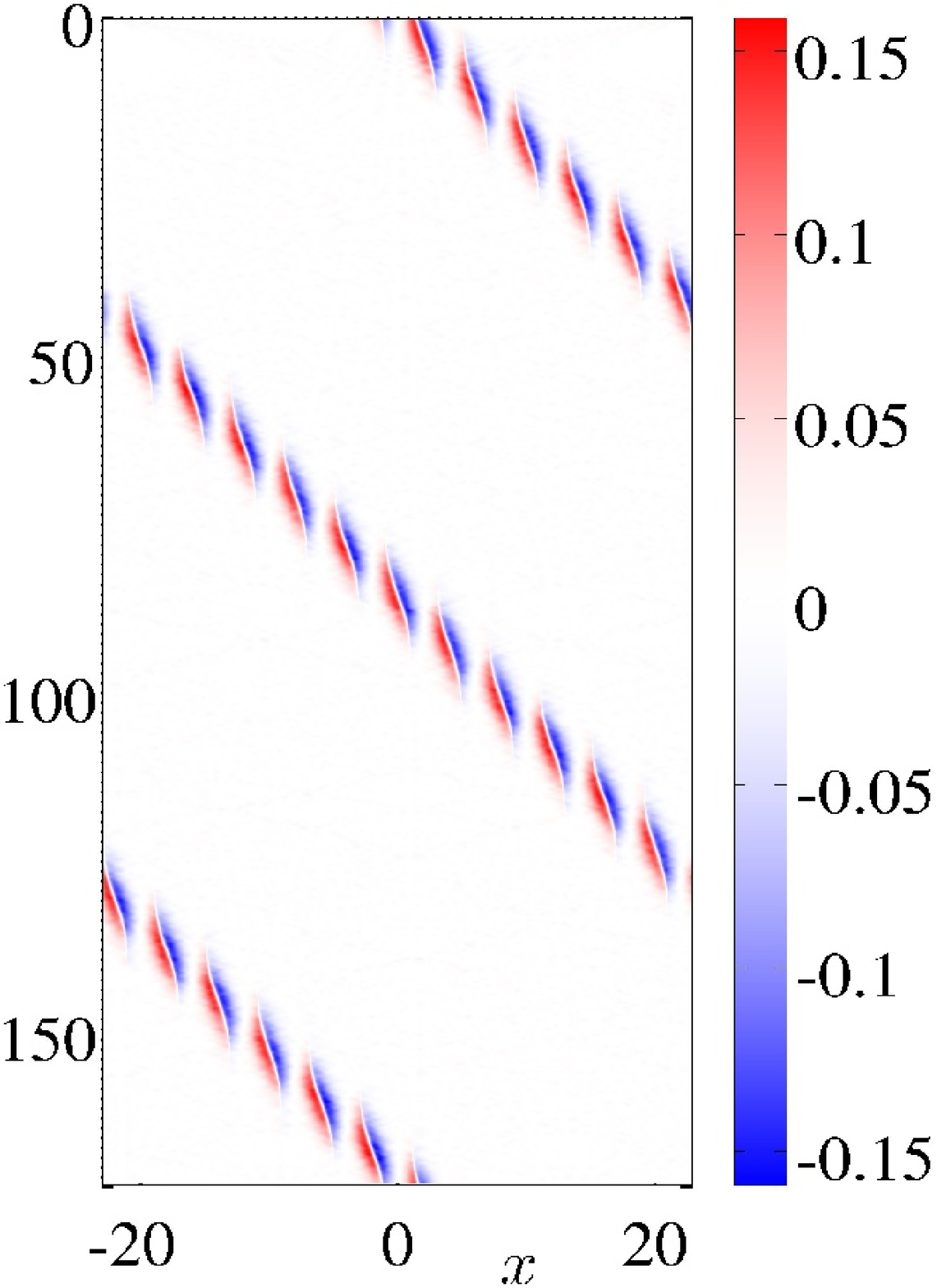,height=4.4cm,silent=} 
\end{center}
\caption{{Color online}.
Toda lattice soliton riding in a chain of $N=12$ DSs on a
periodic domain. The top two rows of panels depict the
initial conditions in terms of the relative displacement from equilibrium
of the mutual distances $r_n(0)$ (top subpanels) and their corresponding
velocities (bottom subpanels).
The middle and bottom row of panels depict, respectively, the evolution
of the (square root of the) density $|u(x,t)|$ and its time derivative
for the original GP model initialized with the Toda lattice soliton.
The left, middle and right columns correspond to increasingly large
Toda lattice velocities for a background density $n_0=1$ and, 
from left to right,
$r_0=6$, $r_0=5$, and $r_0=4$, which correspond, respectively, to
$c=0.0180$, $c=0.0489$, and $c=0.1329$.
The dashed lines correspond to the reduced ODE model of the Toda lattice
(\ref{eq:DS_ODE}).
}
\label{fig:u2_singles}
\end{figure}

\begin{figure}[t] 
\begin{center}
\hskip-0.95cm
\psfig{figure=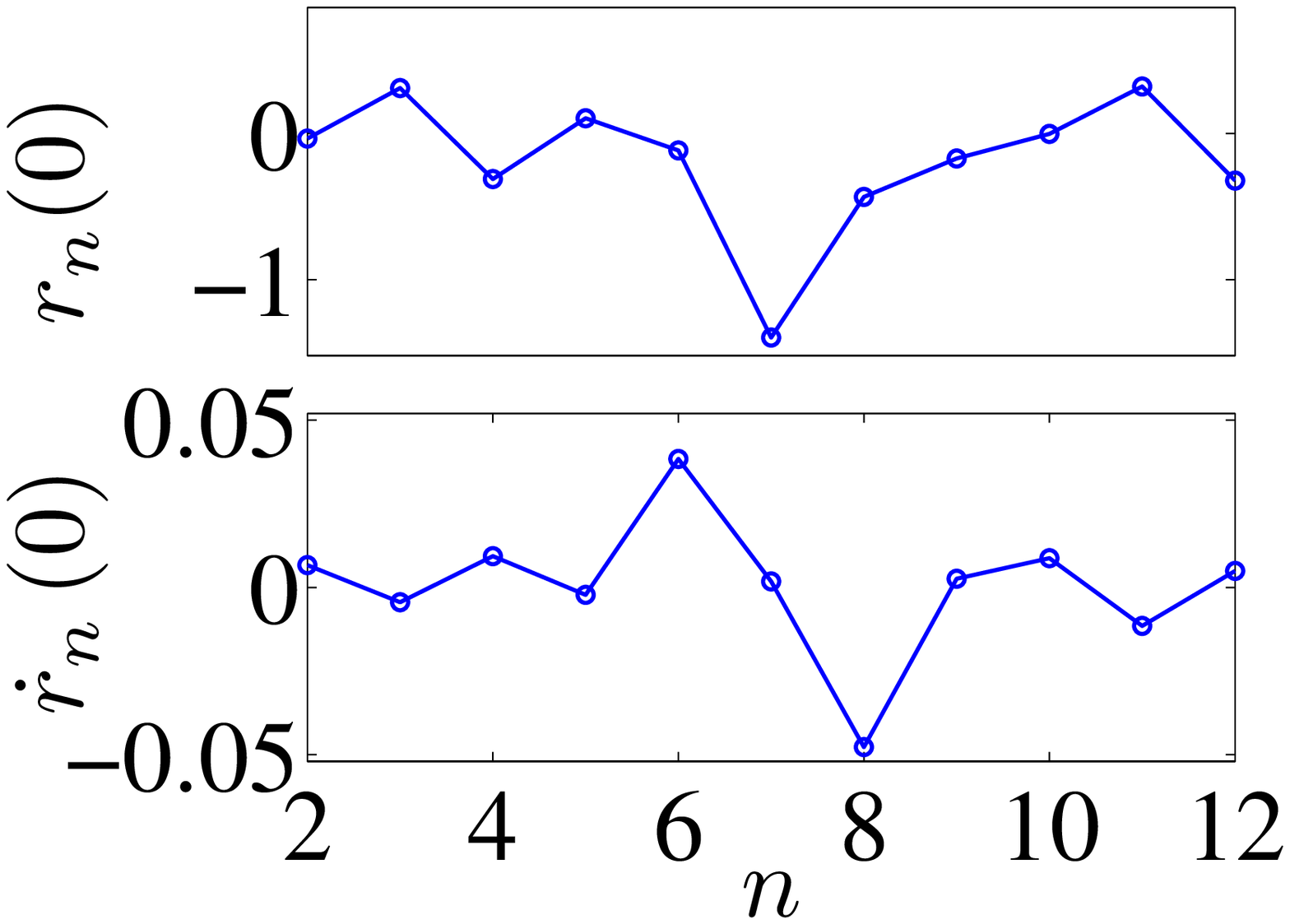,height=2.10cm,silent=} 
\hskip-0.15cm
\psfig{figure=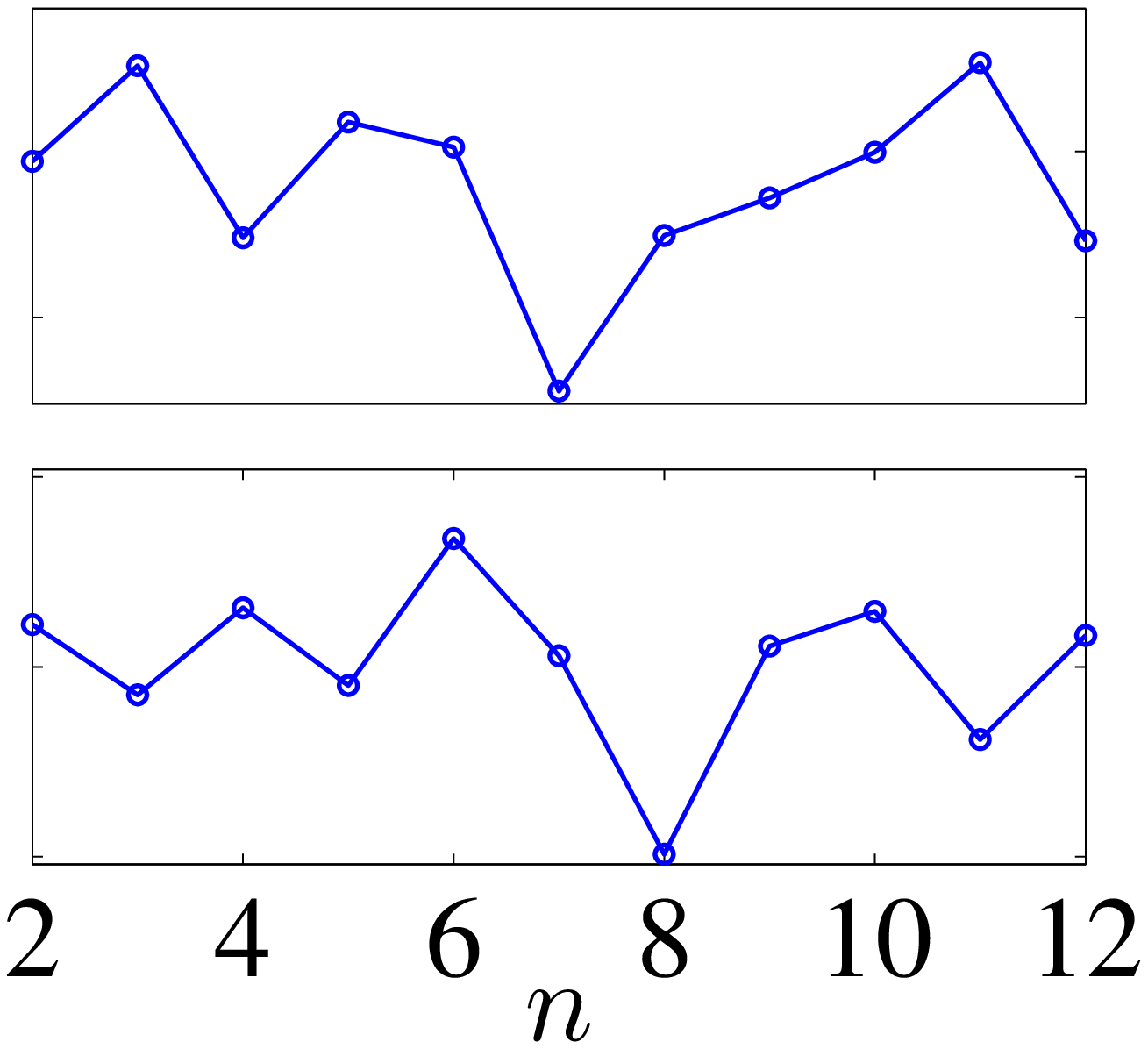,height=2.10cm,silent=} 
\hskip-0.15cm
\psfig{figure=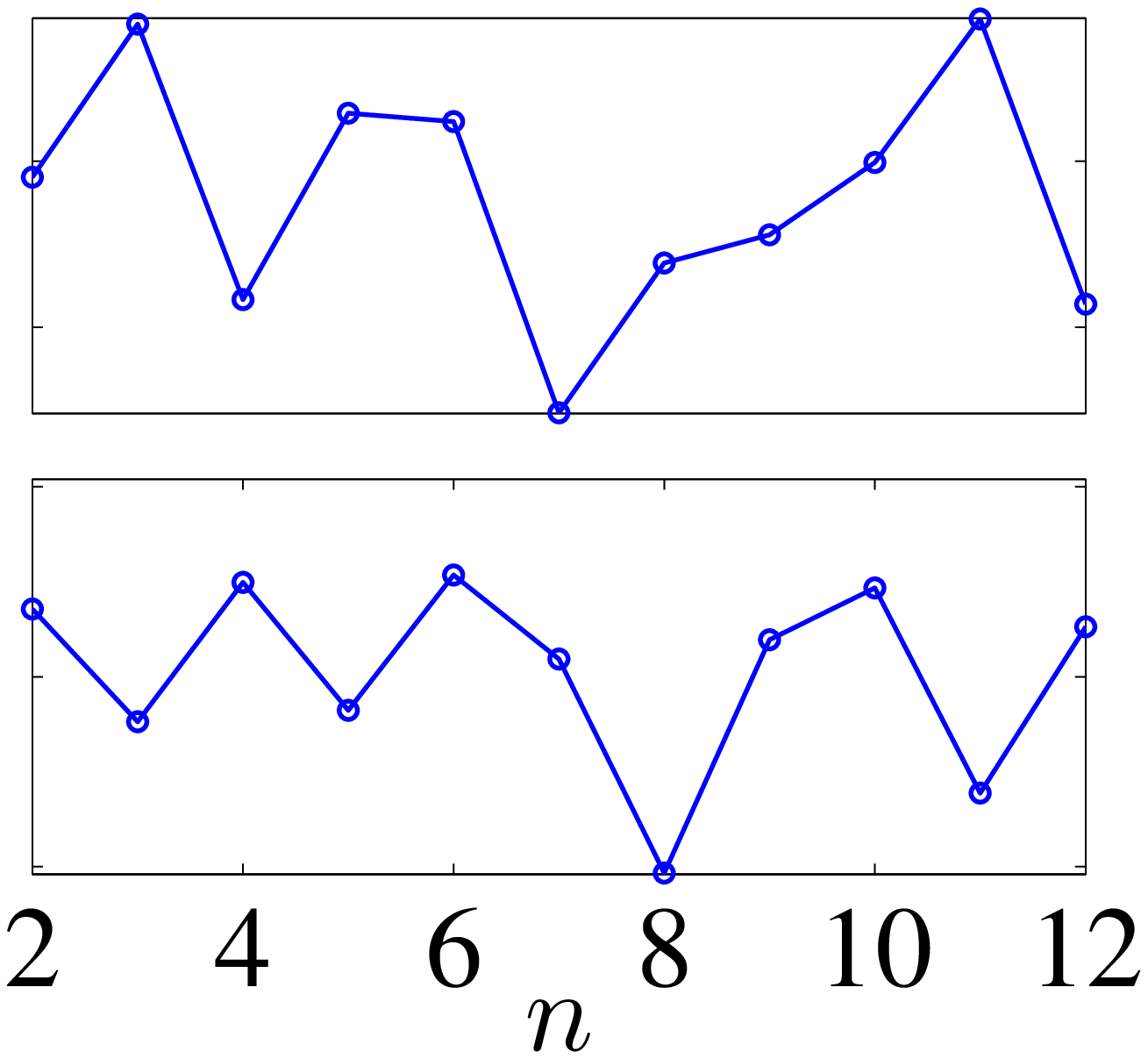,height=2.10cm,silent=} 
\\
\hskip-0.20cm
\psfig{figure=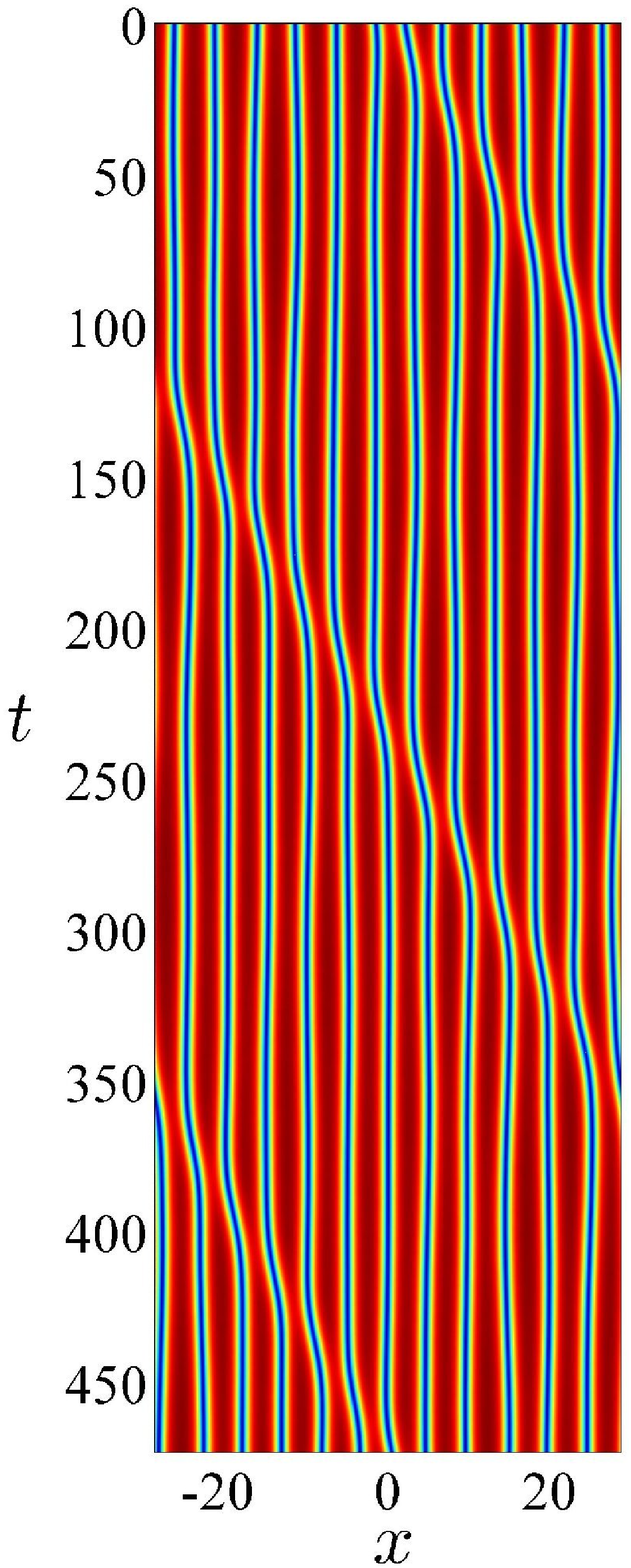,height=7.3cm,silent=} 
\psfig{figure=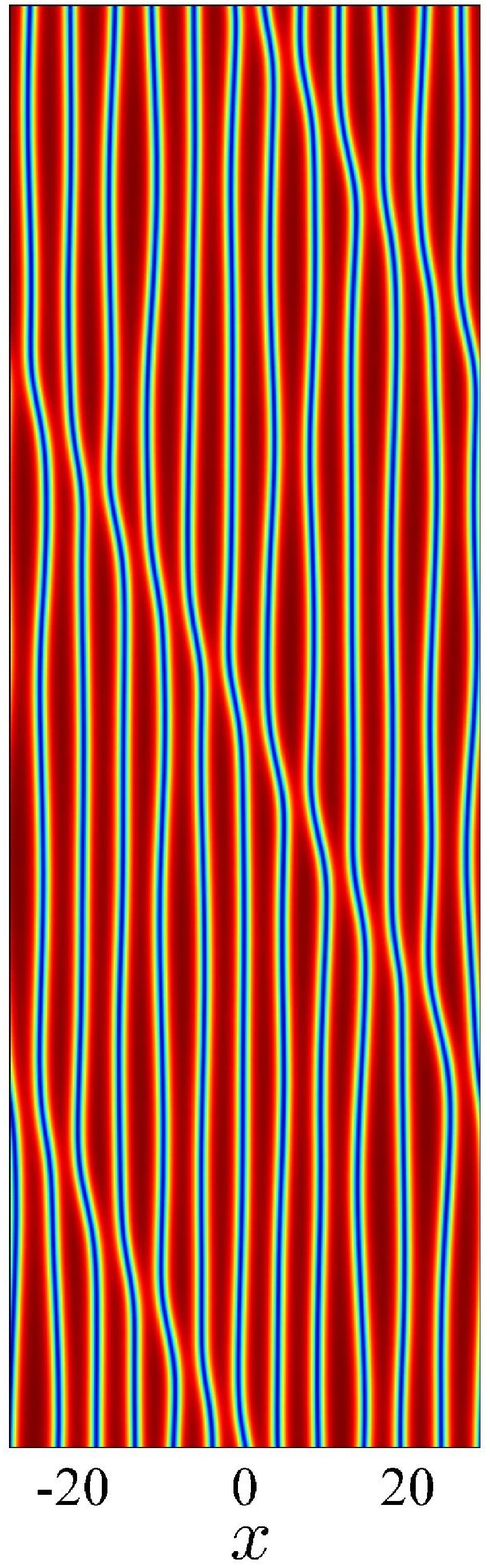,height=7.3cm,silent=} 
\psfig{figure=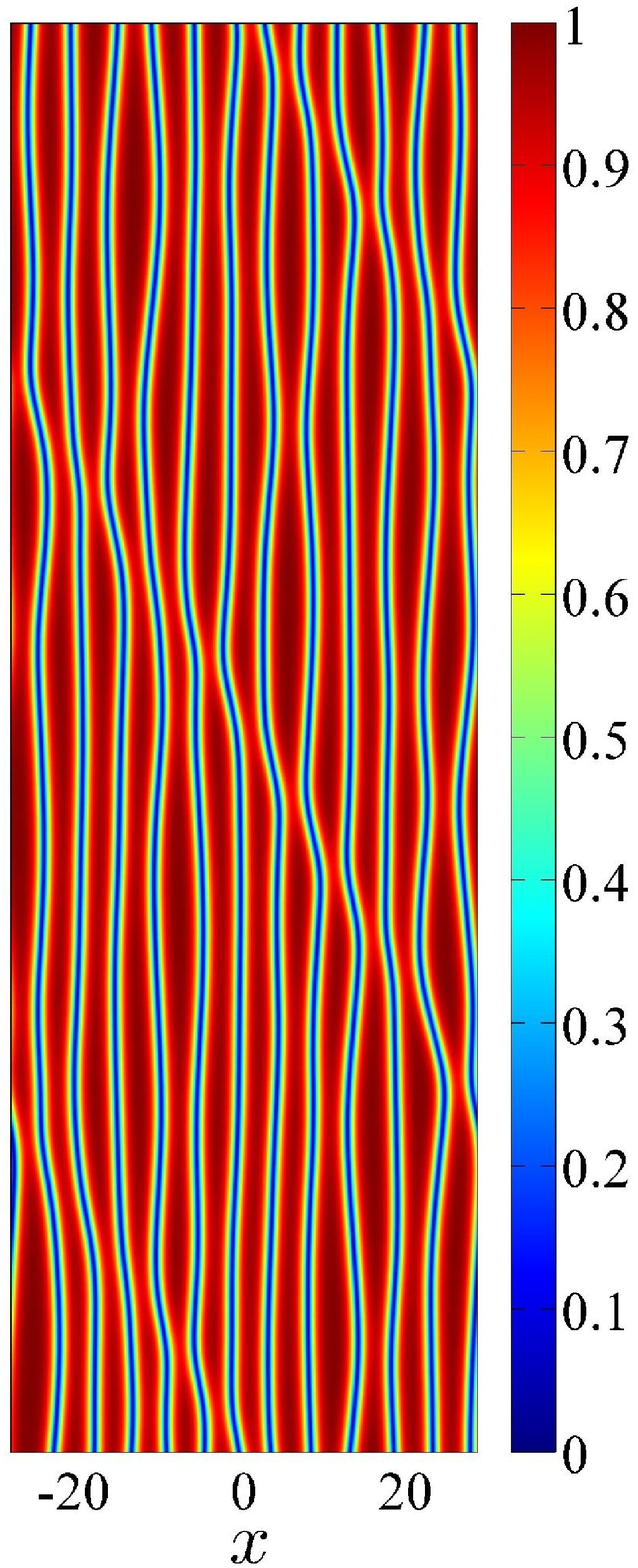,height=7.3cm,silent=} 
\\
\psfig{figure=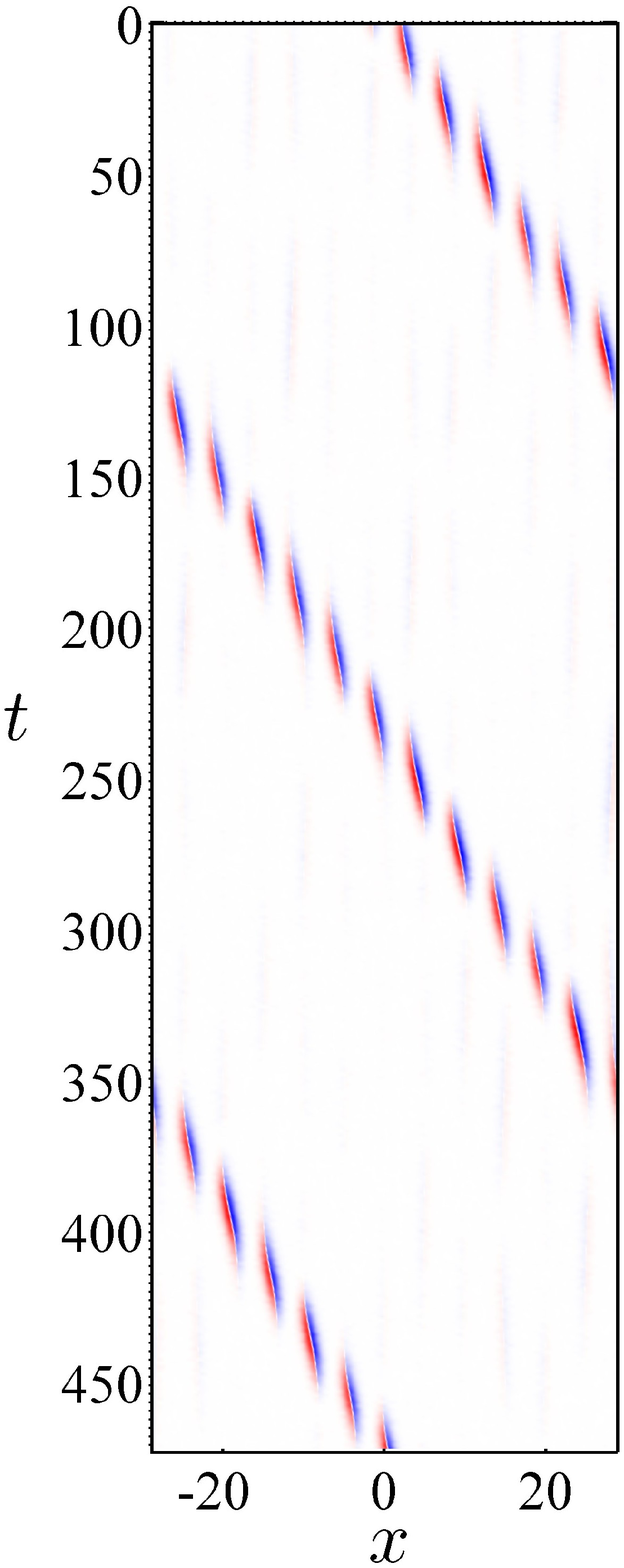,height=7.3cm,silent=} 
\psfig{figure=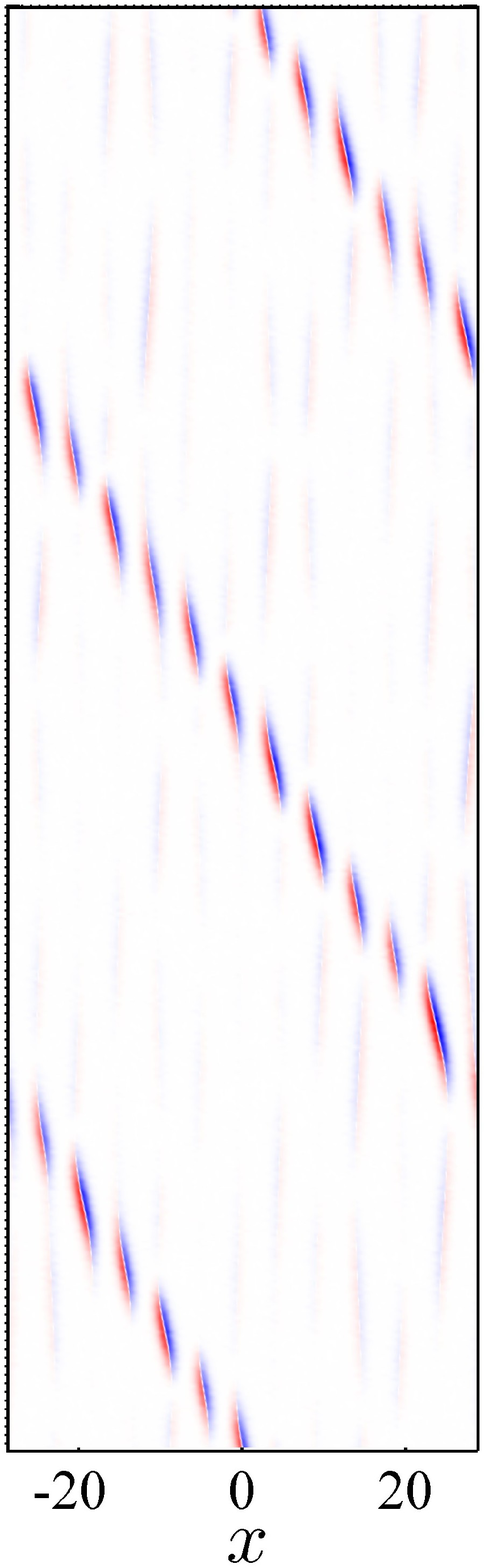,height=7.3cm,silent=} 
\psfig{figure=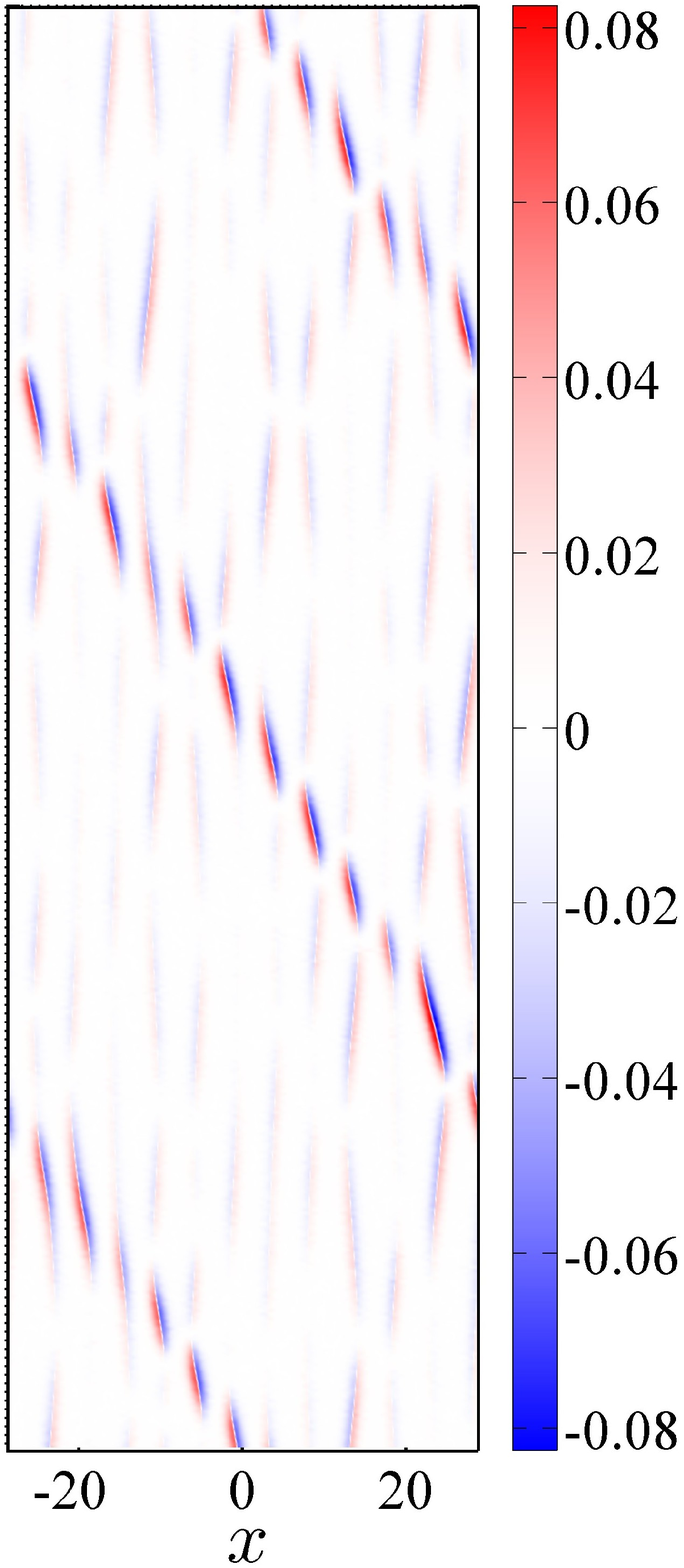,height=7.3cm,silent=} 
\end{center}
\caption{{Color online}.
Stability of the hypersoliton. The layout of the different panels is the
same as in Fig.~\ref{fig:u2_singles}. The initial condition corresponds to the
case of the middle column of panels of Fig.~\ref{fig:u2_singles} ($r_0=5$)
with random
perturbations added to the initial condition. The initial condition is
perturbed by adding a random perturbation of size, from left to right, 30\%,
50\%, and 80\% on the initial separations and velocities of the DSs in
the chain.
}
\label{fig:u2_perturb}
\end{figure}

Let us now turn our attention to the DS case. 
Here, the problem of the desynchronization
of the phases is naturally avoided and thus it should be possible to observe
Toda solitons travelling stably through the DS chain.
In fact, this is precisely what we observe in our numerics for a wide range
of parameters for the DSs themselves {\em and} of the Toda soliton.
Figure~\ref{fig:u2_singles} depicts three of such examples corresponding to
three difference Toda soliton initial velocities. The different
velocities are computed from the Toda lattice soliton solution
(\ref{eq:TL_kink}) corresponding to the reduced Toda model for the DSs
(\ref{eq:DS_ODE}) for $n_0=1$ and unperturbed initial separations
$r_0=6$, 5, and 4.
The top panels in the figure depict the initial compression wave ($r_n$) 
and its initial velocity ($\dot r_n$). 
The middle row of panels depicts the dynamics on the (square root of
the) density ($|u|$) together with the reduced ODE Toda model (\ref{eq:DS_ODE})
superimposed to it.
The bottom row of panels depicts the time derivative of the (square root of
the) density ($|u|_t$) illustrating that outside of the region of localization of the Toda
soliton, there are no perturbations indicating a stable, radiationless,
propagation of the Toda soliton. From now on, we dub such a solution a {\em hypersoliton}
as it is a (Toda lattice) soliton riding on a chain of (dark) solitons.
As shown in Fig.~\ref{fig:u2_singles}, the reduced Toda lattice and, in particular, its
Toda lattice solution, accurately describes the behavior of the original GP model.

In order to further study the stability and robustness of the crafted
hypersolitons, we proceed to initialize the lattice with an initial 
condition corresponding to a perturbed hypersoliton. 
Specifically, Fig.~\ref{fig:u2_perturb} shows the dynamical evolution 
after using 30\%, 50\%, and 80\% random initial
perturbations to, both, initial relative positions and their
respective velocities.
It is remarkable that the addition of such high levels of perturbations
to the initial conditions do not destroy the hypersoliton. Instead, these
perturbation seem to just provide some background noise over which the
hypersoliton rides with minimal interaction.
This background noise is more clearly visible in the bottom row of
panels depicting the time derivative of the (square root of the) density.
This remarkable robustness of the hypersoliton even after the addition of such
large perturbations to the initial conditions ---that in turn develop 
into perturbations along the whole domain--- is due to two independent
facts: (a) on the one hand, DSs for the GP equation are very robust,
a property owing from their topological charge, and
(b) the structural stability of Toda solitons of the Toda lattice.
These two stability properties, at two distinct levels of the model,
combine to give the hypersoliton on the original GP model its 
robustness.

\begin{figure}[t] 
\begin{center}
\hskip-0.95cm
\psfig{figure=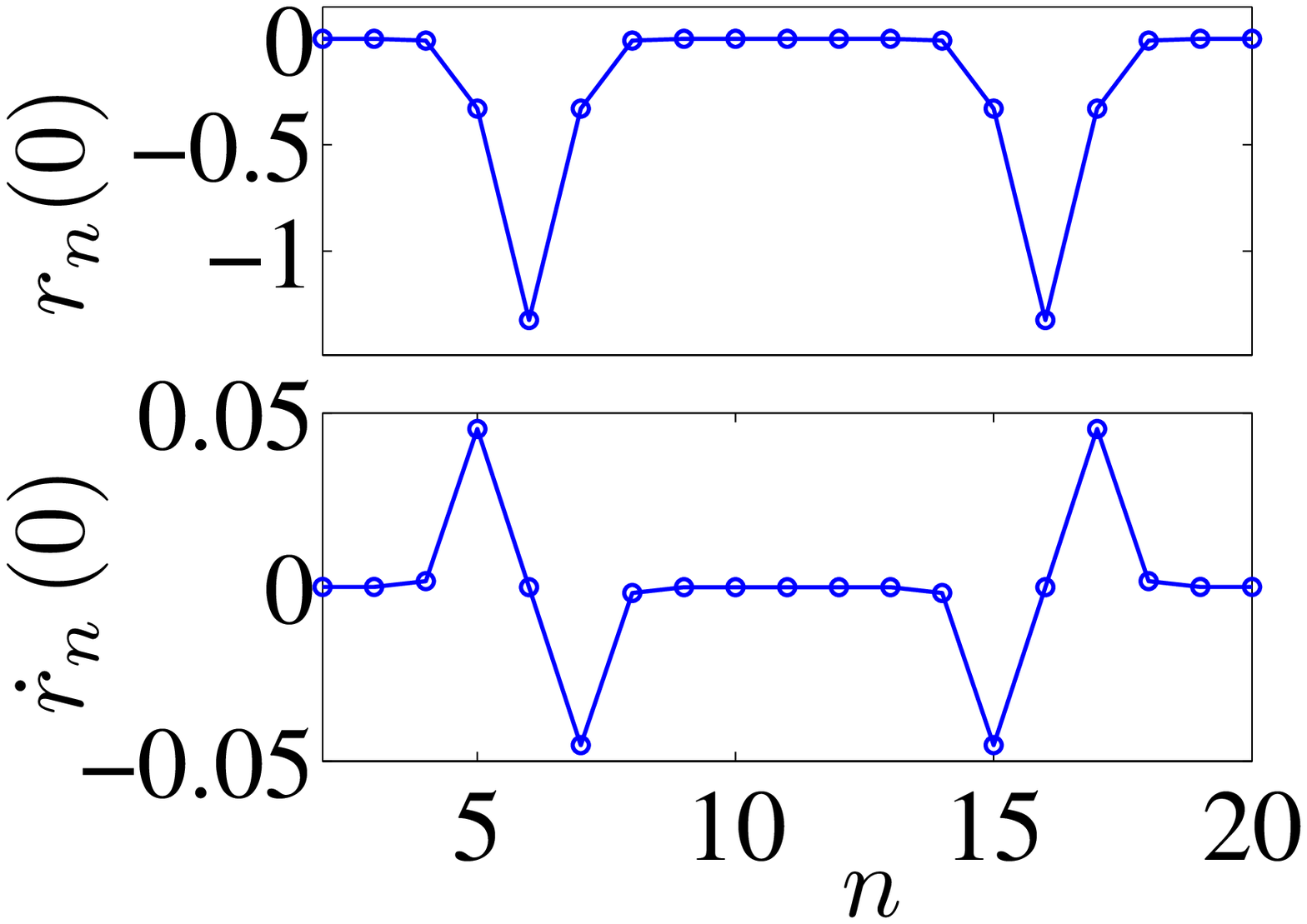,    height=2.1cm,silent=} 
\hskip-0.25cm
\psfig{figure=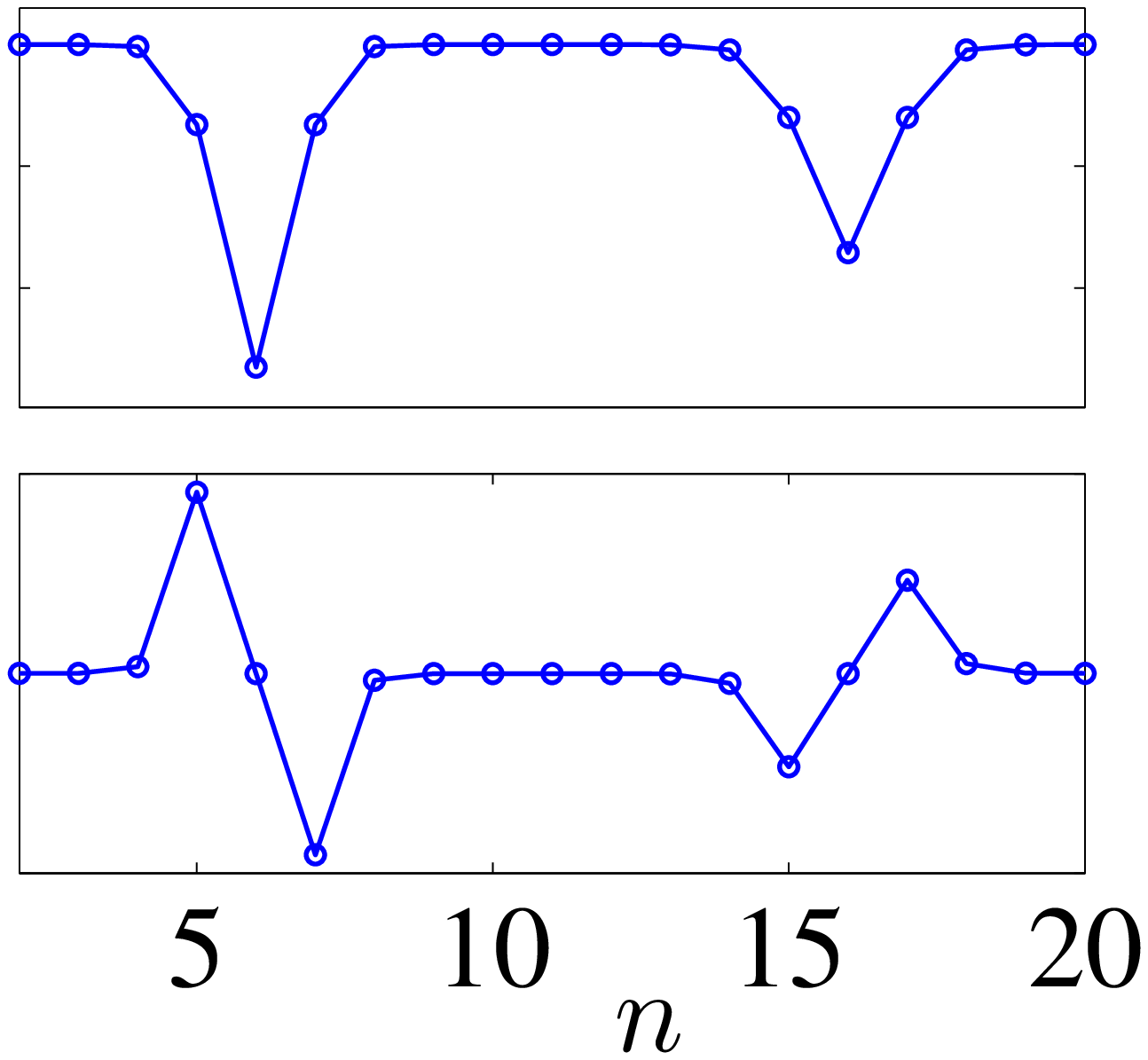,height=2.1cm,silent=} 
\hskip-0.25cm
\psfig{figure=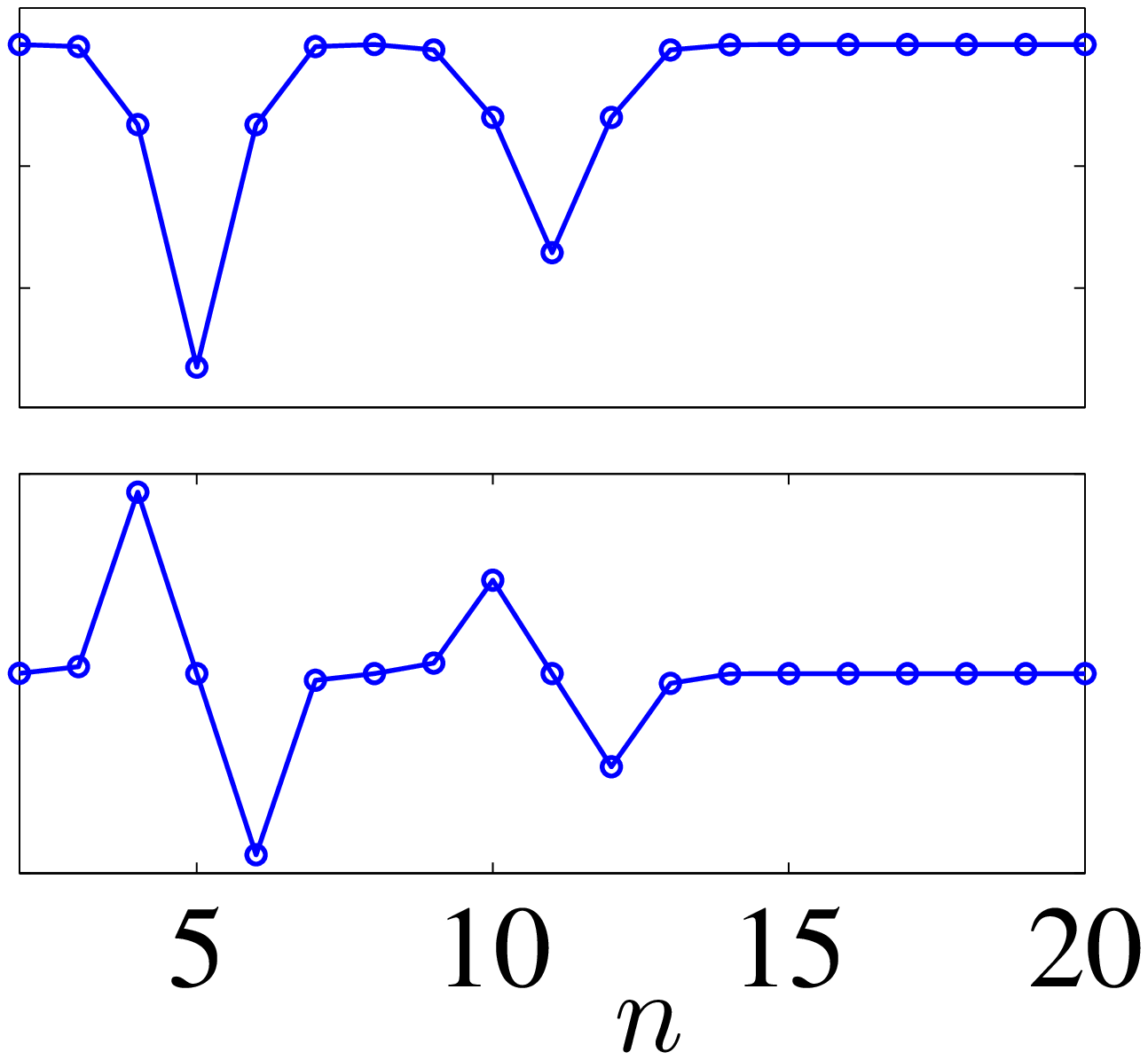,       height=2.1cm,silent=} 
\\
\hskip-0.2cm
\psfig{figure=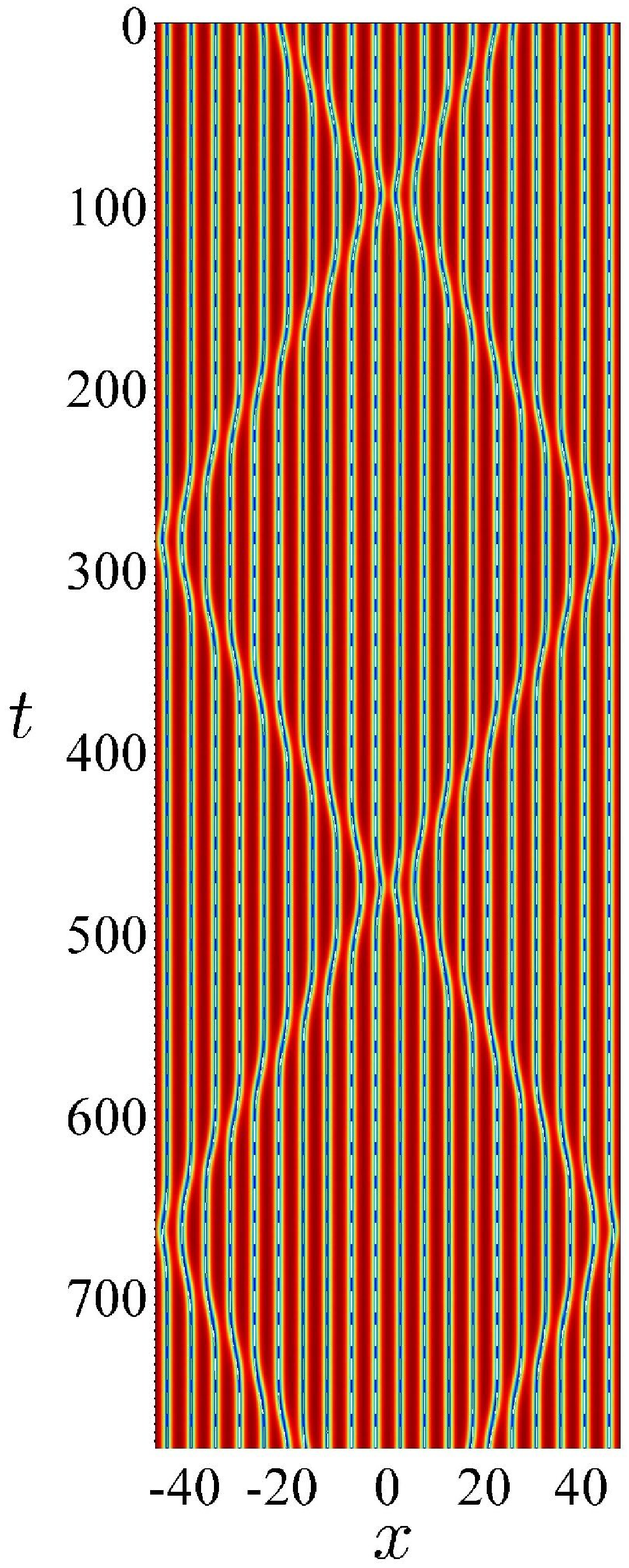,    height=7.3cm,silent=} 
\psfig{figure=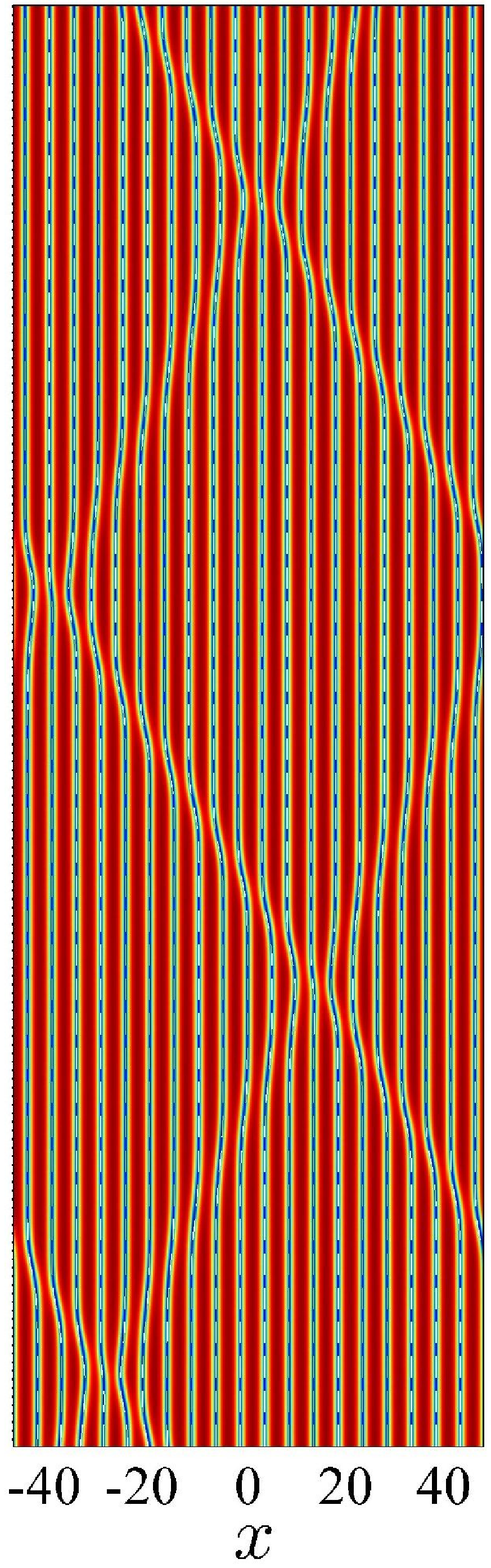,height=7.3cm,silent=} 
\psfig{figure=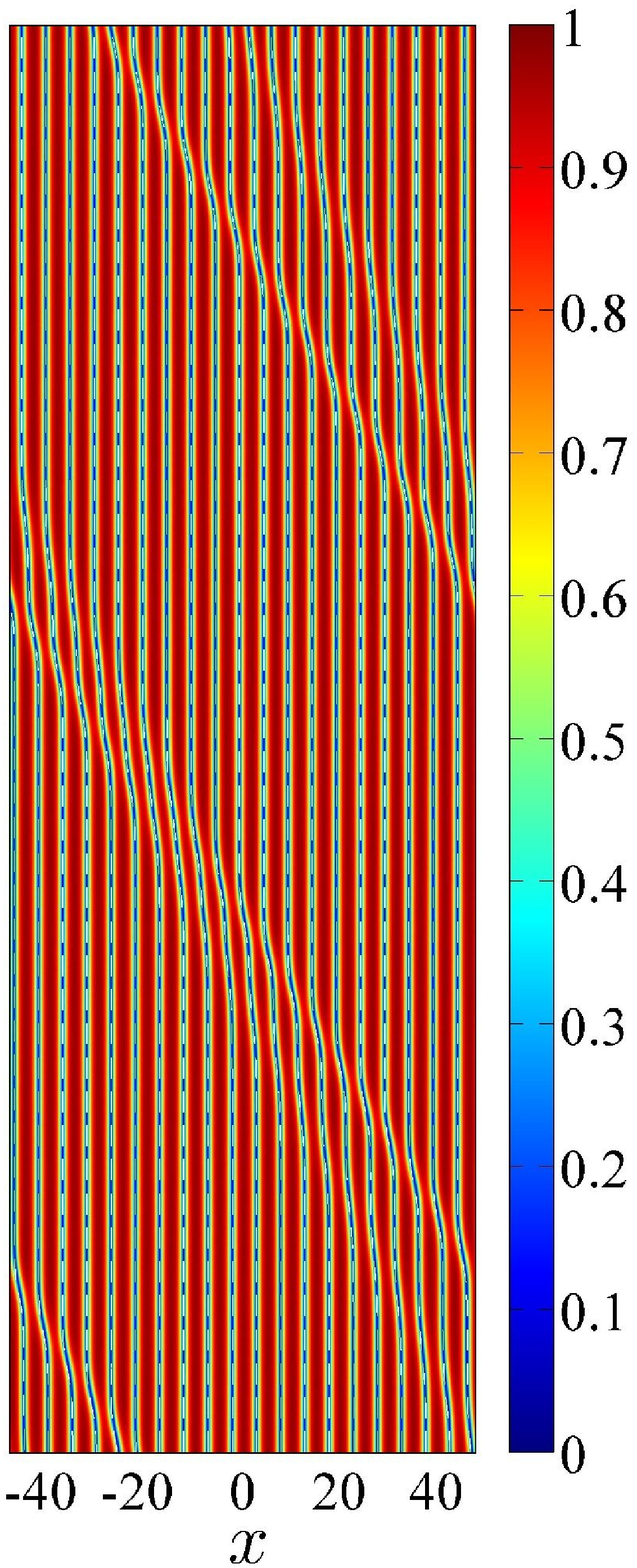,       height=7.3cm,silent=} 
\\
\psfig{figure=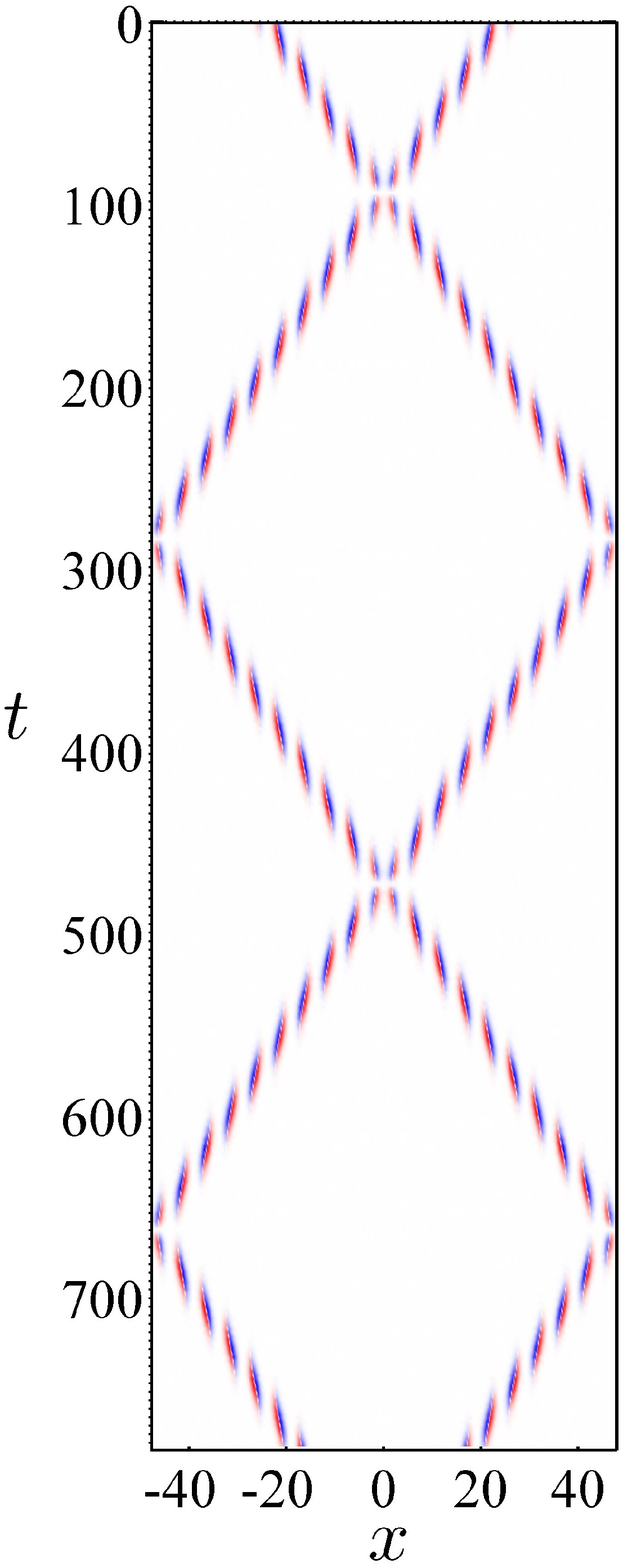,    height=7.3cm,silent=} 
\psfig{figure=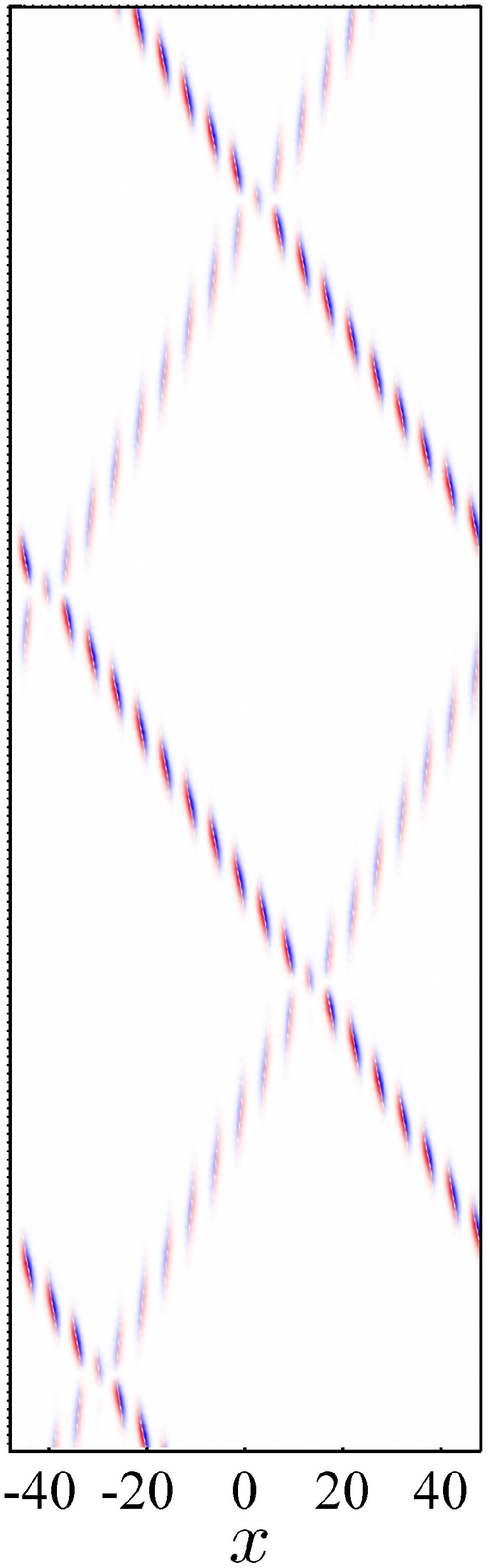,height=7.3cm,silent=} 
\psfig{figure=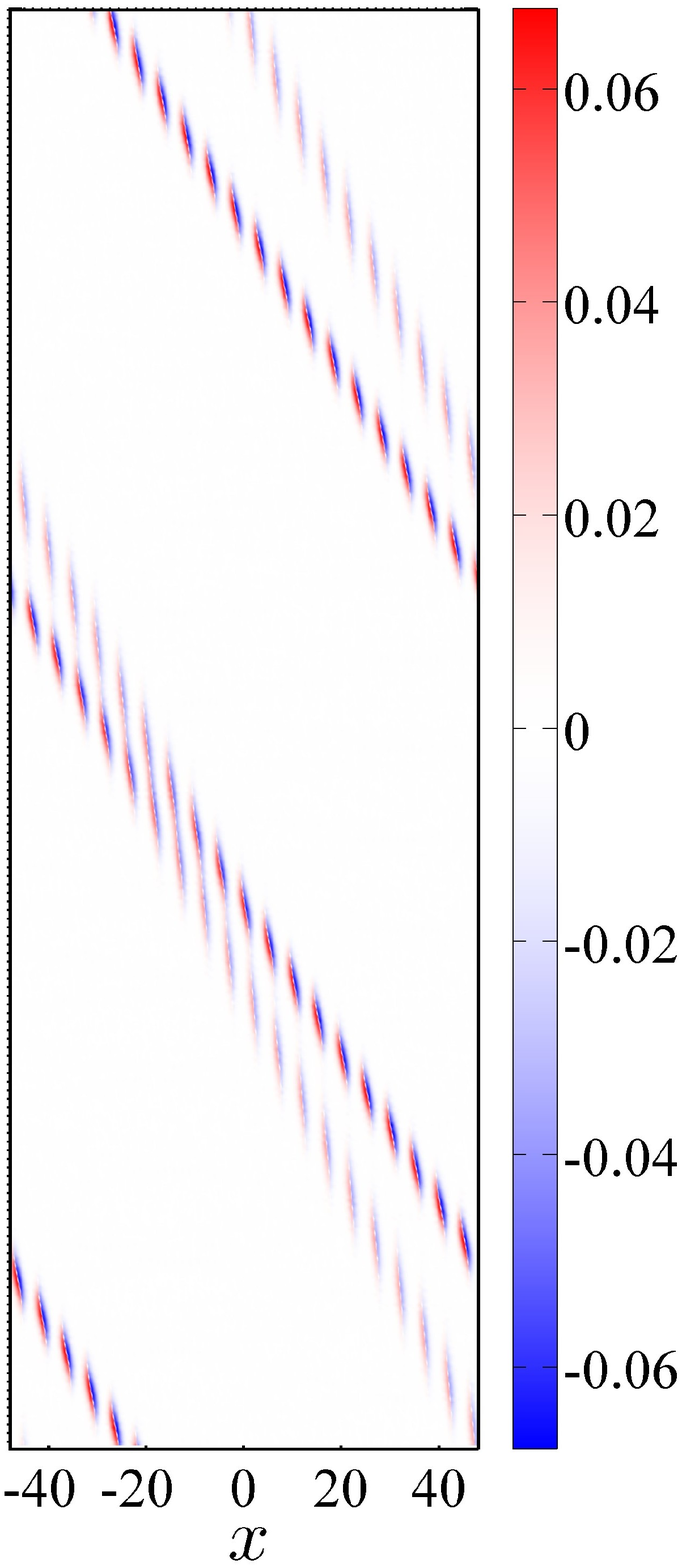,       height=7.3cm,silent=} 
\end{center}
\vspace{-0.2cm}
\caption{{Color online}.
Hypersoliton collisions. The left, middle and right columns depict,
respectively, the collision for
(a) a head-on collision
with equal but opposite velocities,
(b) a head-on collision
with different speeds,
and
(c) a chasing collision where a faster hypersoliton (seeded on the
left) chases a slower hypersoliton (seeded on the right)
until they collide in a nonlinear fashion.
All panels have the same meaning and layout as previous figures.}
\label{fig:u2_collisions}
\end{figure}

\subsection{Toda lattice solitons: hypersoliton collisions}

Now that we have established the existence and stability of the
hypersoliton solutions in DS chains, it is possible to study several
dynamical aspects arising at this higher structural level.
For example, one can initialize the DS chain with two (or more)
hypersolitons and allow them to interact. It is expected that the
dynamics governing the interactions between hypersolitons will be 
prescribed by the corresponding dynamics on the Toda lattice.
In particular, Toda lattice solitons collide elastically without
energy loss during the collision process.
This is precisely what we observe for a wide range of cases when
seeding two hypersolitons at different locations with different
initial speeds (if they have the same velocity they will chase each
other indefinitely).
Figure \ref{fig:u2_collisions} shows typical examples of hypersoliton
collisions. Specifically, the left, middle and right cases correspond,
respectively to:
(a) a head-on collision for opposite velocities with the same magnitude,
(b) a head-on collision for velocities with different magnitudes,
and
(c) two chasing hypersolitons where a faster one chases a slower one
until they collide.
As the bottom panels show, all these collisions, as expected, are
elastic (i.e., no energy is lost from the travelling hypersolitons
before or after the collisions).
It is interesting to note, as per the ``particle'' nature of the hypersoliton
structures, owed to their nonlinear character, there is a shift in their
paths when comparing their straight line trajectories [in the $(x,t)$ plane]
before and after collision.
This effect is clearly visible in the last case of Fig.~\ref{fig:u2_collisions}
where the fast soliton is advanced after collision while the slow one is
delayed after collision.
This effect will be important as we construct a quantum analog of the classical
Newton's cradle using hypersolitons in a chain of DS trapped inside a
confining potential in the next section.

\subsection{Quantum Newton's cradle}
\label{sec:NewtonCradle}

We now proceed to study the effects of adding a confining external
potential [i.e., $V_{\rm MT} \not=0$ in Eq.~(\ref{GPE})],
relevant to the modelling of magnetically trapped BECs, on the dynamics
of hypersolitons supported by a finite DS chain.
In the presence of the external parabolic potential (\ref{eq:VMT}), a single DS
exhibits approximately harmonic oscillations with a frequency
$\omega=\Omega/\sqrt{2}$ (see Refs.~\cite{DJF_REVIEW,DARK_BOOK,pelpan,fr4,fr5}
and references therein). This result is valid in the so-called Thomas-Fermi limit
corresponding to the high density limit.
Therefore, combining ---through perturbation theory---
the mutual interactions between DSs and the force exerted by the
external trap yields~\cite{DJF_REVIEW,DARK_BOOK}
\begin{equation}
\ddot\xi_i = 8n_0^{3/2}\,e^{-2\sqrt{n_0}(\xi_{i}  -\xi_{i-1})}
           - 8n_0^{3/2}\,e^{-2\sqrt{n_0}(\xi_{i+1}-\xi_{i})}
           - \omega^2 \xi_i,
\label{eq:DS_ODE_MT}
\end{equation}
corresponding to the Toda lattice described by Eq.~(\ref{eq:DS_ODE})
with an added on-site force generated by the external potential
on each of the DSs of the chain.
The above model has not only been validated numerically, but it
has been used to predict the normal modes of vibration for a small
number of DSs in actual experiments~\cite{kip,kip2}.
In fact, the Toda lattice with the on-site potential (\ref{eq:DS_ODE_MT})
possesses a steady state solution emerging from the balance of the
mutual repulsions between the DSs and the attraction of the external
trap towards the trap's center.
This compressed steady state train of $N$ DSs has $N$ distinct normal 
modes of vibration corresponding to the normal modes of
vibration of $N$ coupled masses through (linear) springs with springs
at each end attached to rigid walls.
For example, for $N=2$ there exist 2 normal modes of vibration corresponding
to the in-phase and out-of-phase modes of vibration of the DSs~\cite{kip2}.

\begin{figure}[t] 
\begin{center}
\hskip-0.5cm
~~
\psfig{figure=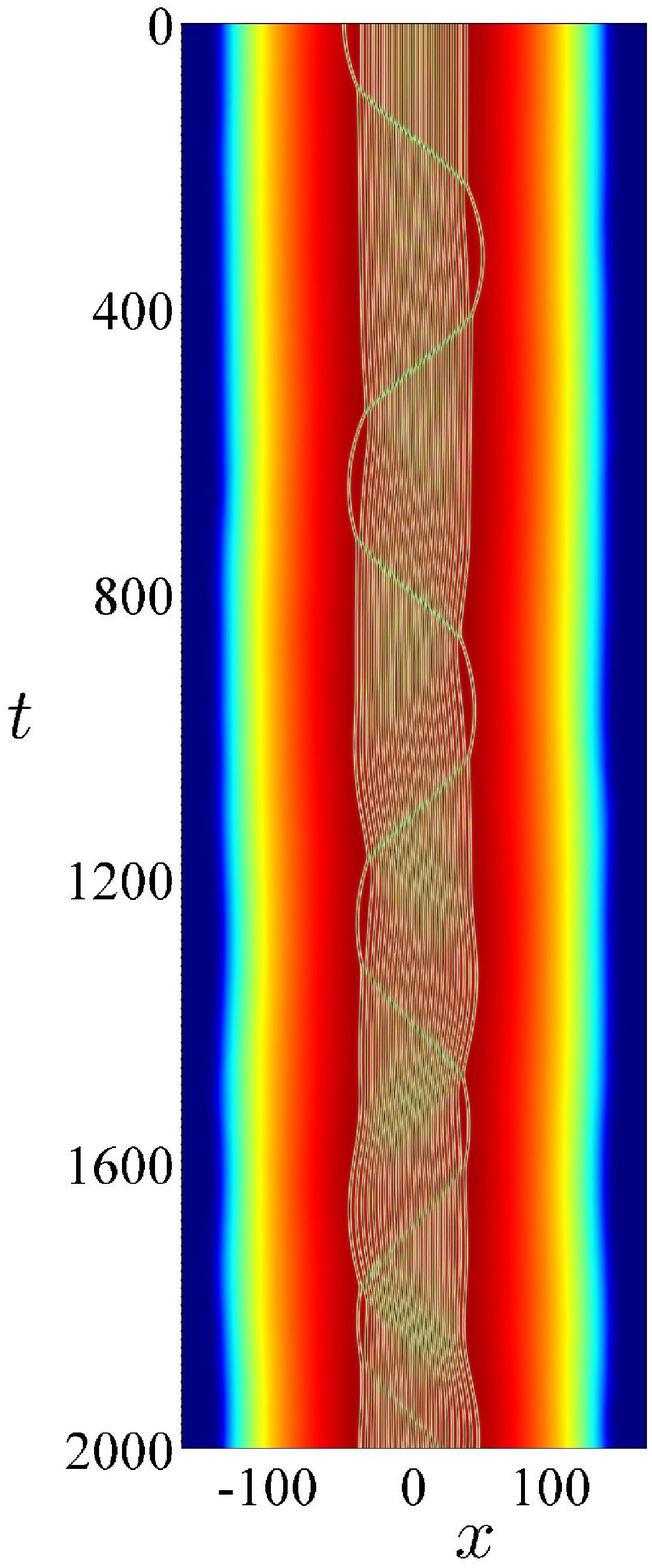,height=7cm,silent=} 
\psfig{figure=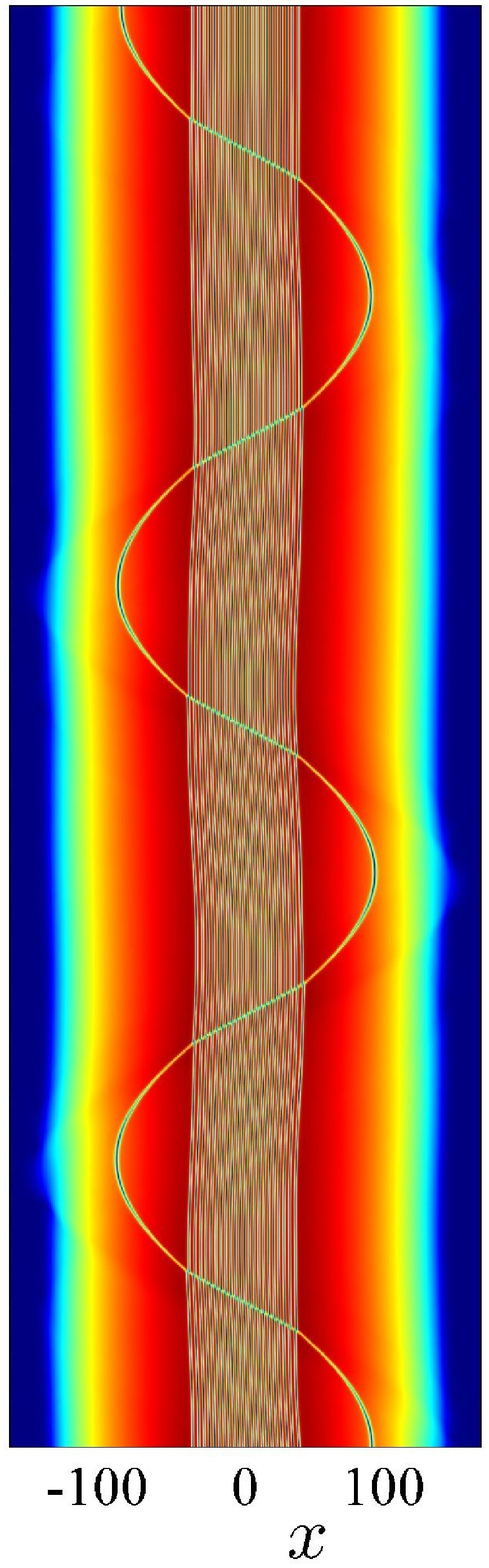,height=7cm,silent=} 
\psfig{figure=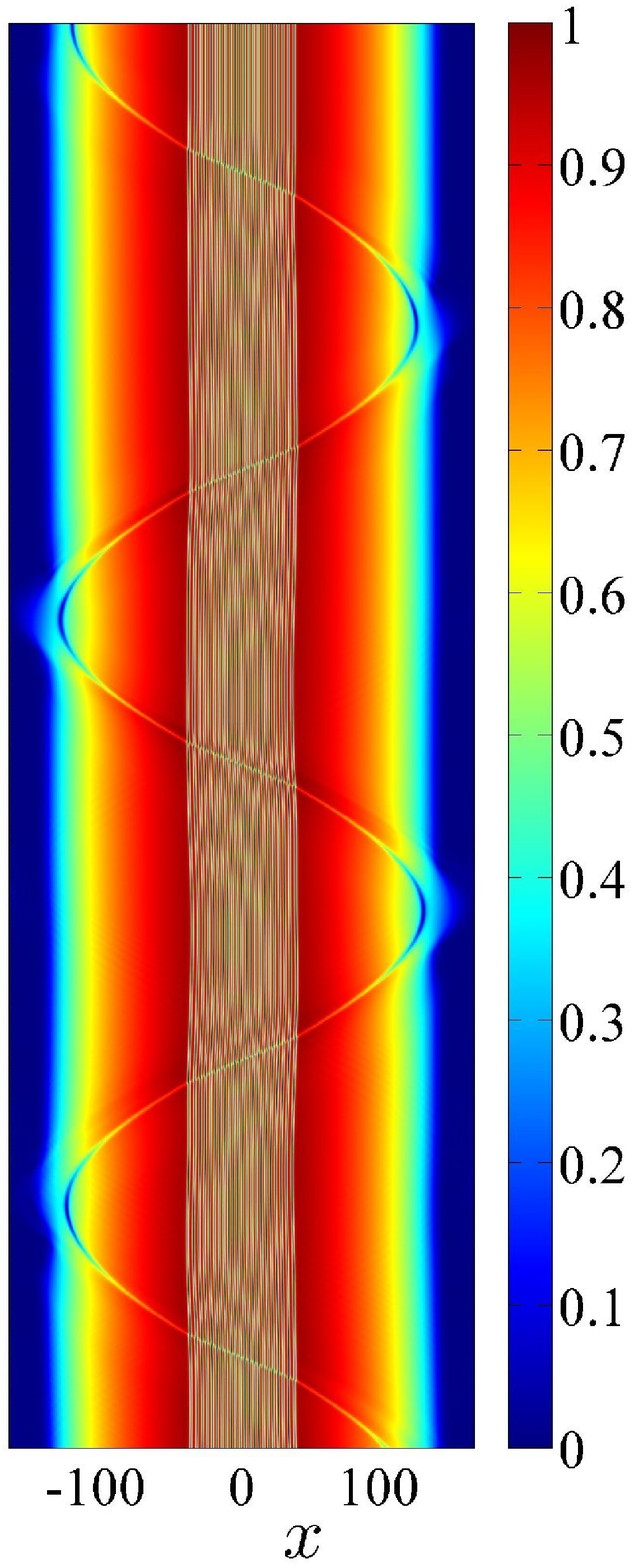,height=7cm,silent=} 
\\
\psfig{figure=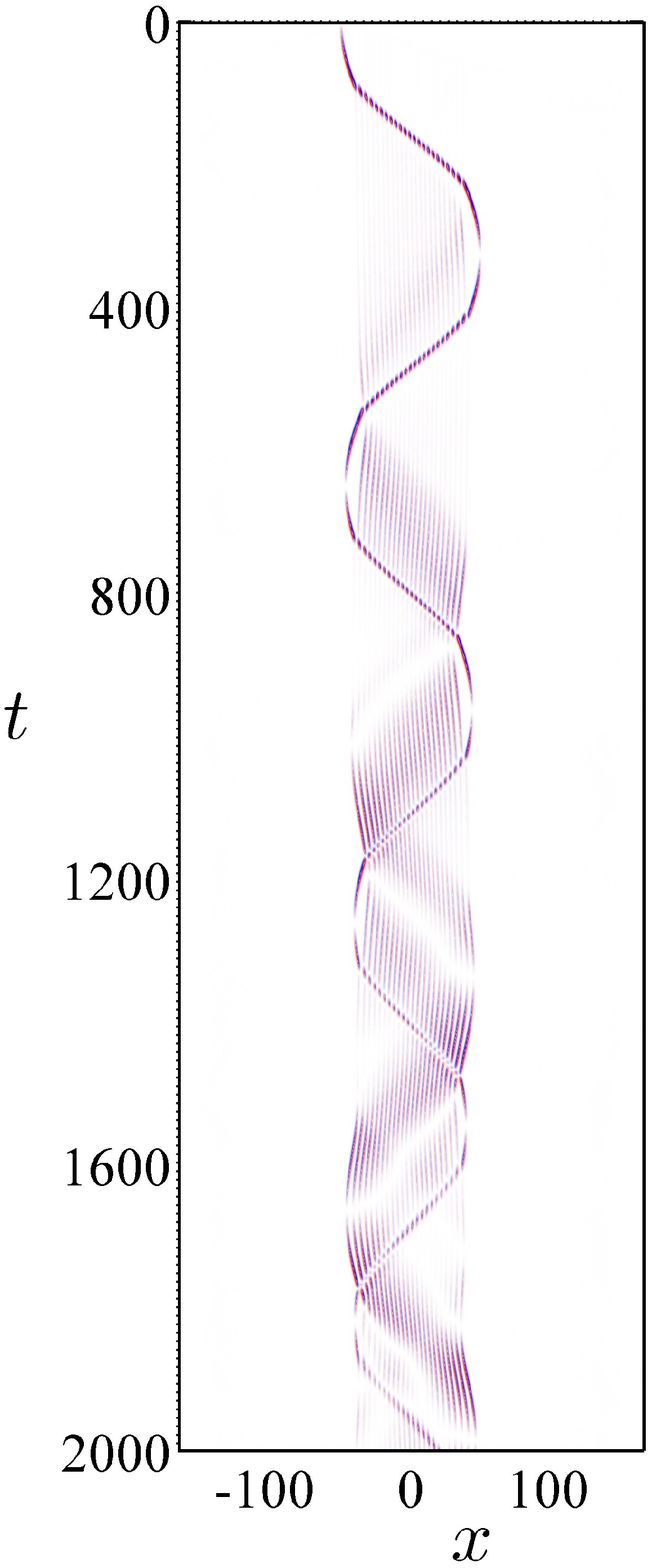,height=7cm,silent=} 
\psfig{figure=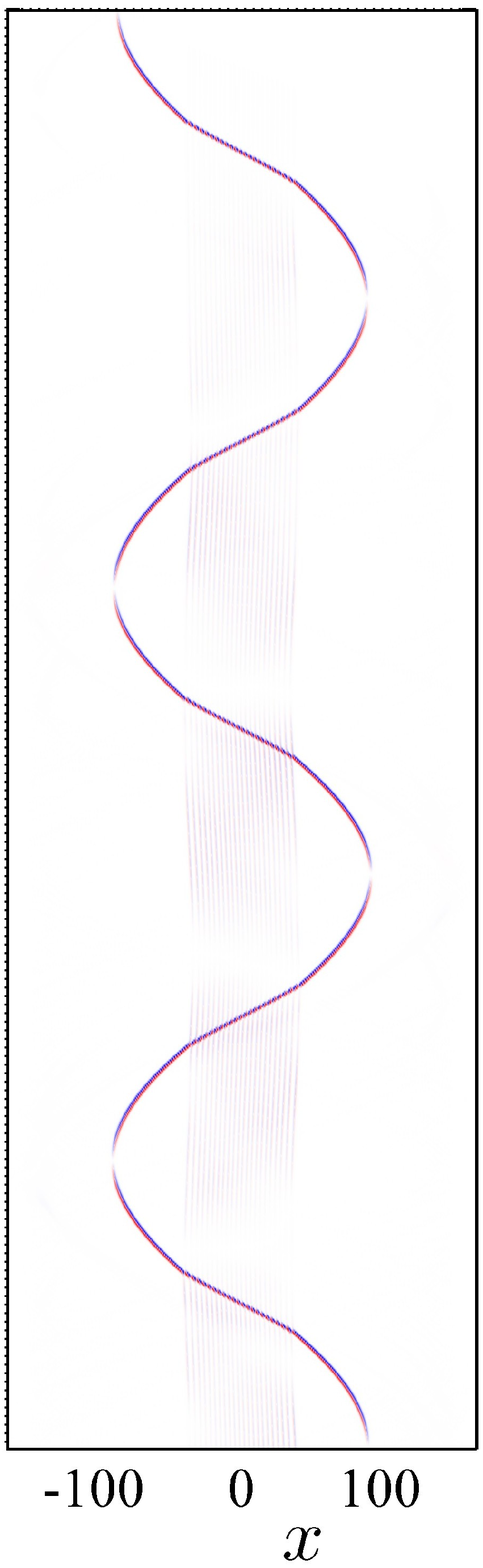,height=7cm,silent=} 
\psfig{figure=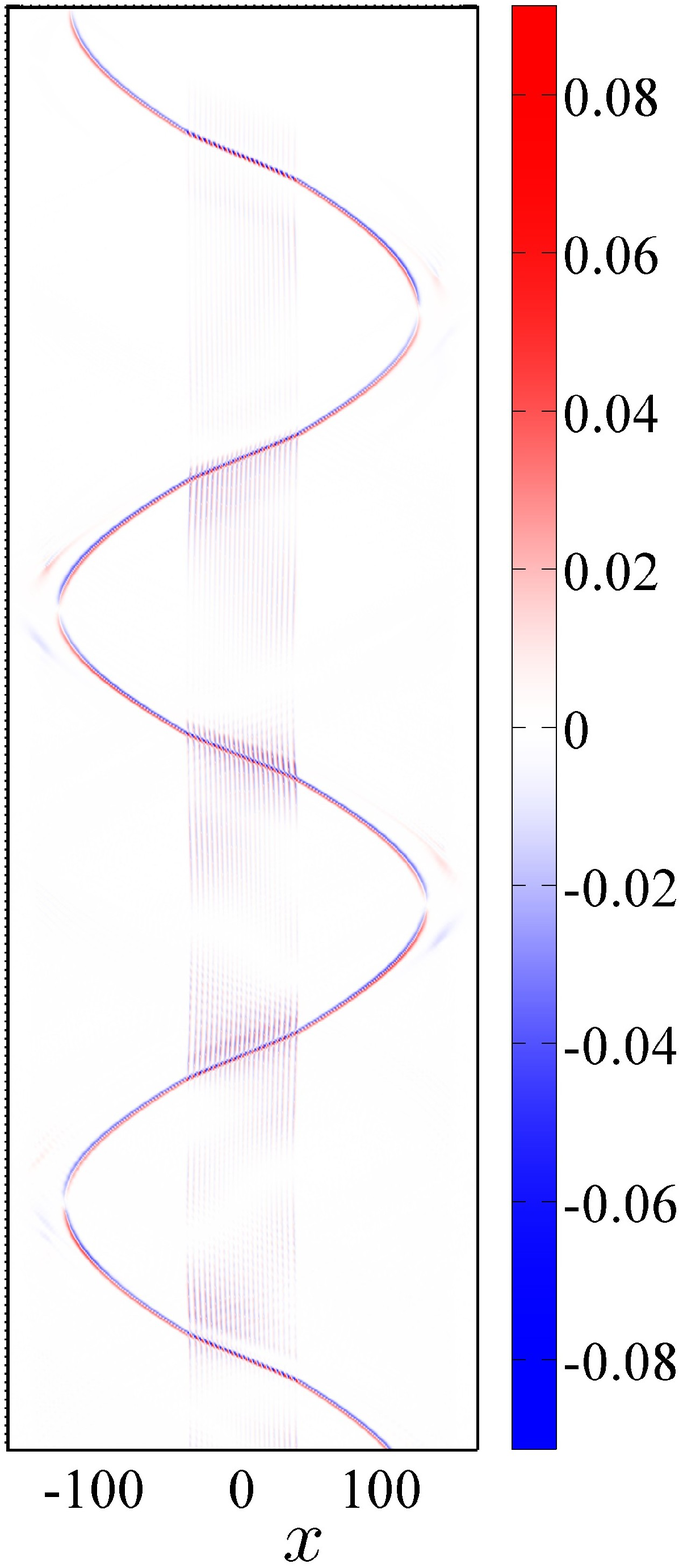,height=7cm,silent=} 
\end{center}
\caption{{Color online}.
Hypersoliton Newton's cradle.
A DS soliton (leftmost one) is released at various distances away from
a stationary lattice of 12 DSs placed at the center of the parabolic trapping
potential (\ref{eq:VMT}) with $\Omega=0.01$.
The DS is released at a distance equivalent to
(a) $3r_0$ (left column of panels),
(b) $12r_0$ (middle column of panels), and
(c) $20r_0$ (right column of panels)
where $r_0$ is the separation between the central DSs.
All panels have the same meaning and layout as previous figures.}
\label{fig:cradle_tf2000}
\end{figure}

\begin{figure}[t] 
\begin{center}
\hskip-0.5cm
\psfig{figure=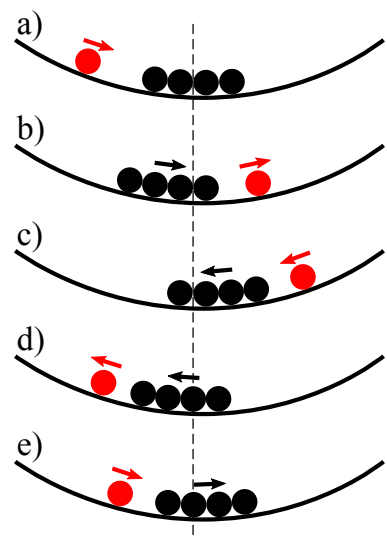,height=7cm,silent=}
\end{center}
\caption{{Color online}.
Schematic of the mechanism for the loss of the
hypersoliton Newton's cradle through the synchronization
of the outer DS (red) with the central DS chain (black).
(a) Release of a DS from the left with a stationary DS chain
in the central region.
(b) After the hypersoliton travels through the chain, the
rightmost DS is ejected to the right. The remaining central DS
chain is now asymmetric with respect to the center (see vertical
dashed line) and starts moving to the right.
(c) Both outer DS and central DS chain perform half an
oscillation after their respective excursion up the trap.
(d) After the DS collides and transfers its energy (through a
propagating hypersoliton through the central DS chain) the leftmost DS
is ejected. The central DS chain still has the extra energy of moving
towards the left.
(e) Both the DS and the central chain now oscillate in the same
direction (although slightly out-of-phase).
This process continues until the energy of the outer DS is
completely transferred to the central chain and all that
remains is a central chain undergoing left-to-right oscillations
in the in-phase normal mode.
}
\label{fig:cradle_schematic}
\end{figure}

Instead of studying further the normal modes mentioned above, 
we opt here to
emulate the dynamics of a classical Newton's cradle using a chain
of solitons.
The idea is to start with an initial stationary chain of $N$ DSs at the
bottom of the parabolic trap and then drop a single, outer, DS from
a position higher up in the trapping potential. This outer DS will experience the
force of the trapping potential and ride down the external trap
to collide with the stationary DS chain.
The collision excites a moving hypersoliton within the inner DS chain. When
the hypersoliton reaches the opposite extreme of the inner DS chain,
a single DS is expelled outward. This new outer DS will ride up and down
the trap until it hits the inner DS chain thus repeating the process
of a soliton analogue of the classical Newton's cradle.
It is relevant to note at this stage that a similar idea
was experimentally achieved in a 1D Bose gas of $^{87}$Rb
atoms by initially splitting a wavepacket into two
wavepackets with opposite velocities. These packets then evolve by
going up and down the trap and periodically colliding
at the center~\cite{Cradle_Nature}.
Also, in Ref.~\cite{Franzosi_14}, the authors propose another
quantum analogue of a Newton's cradle using a BEC by partitioning
the wavefunction using a periodic optical lattice potential.
In contrast, our method to create a Newton's cradle is based on the
effective nonlinearity of the GP model and the repulsive
interaction dynamics between dark solitons.

Figure \ref{fig:cradle_tf2000} depicts three different attempts at
recreating a Newton's cradle-type evolution with our setup. In these examples,
we use a stationary DS chain of $N=12$ DSs placed at the center of
a magnetic trap of strength $\Omega=0.01$, namely $\omega=0.01/\sqrt{2}$
[see Eq.~(\ref{eq:DS_ODE_MT})]. The stationary inner DS chain is obtained
by a fixed-point iteration (Newton's) method initialized with a chain of
DSs with zero velocities positioned at the steady state locations.
Once the stationary inner lattice is found, an extra, outer, DS is
seeded away from it. The distance of the outer DS to the inner DS
chain is varied and the resulting evolution is analyzed.
The three
cases depicted in Fig.~\ref{fig:cradle_tf2000} correspond to, from
left to right, an initial distance of the outer DS of (a) $3r_0$, (b) $12r_0$,
and (c) $20r_0$ where $r_0$ is the distance between the two innermost 
DSs of the central chain.
For case (a), corresponding to a short dropping distance of the DS,
the Newton's cradle dynamics is observed for a couple of periods but
apparently the outer DS is ``absorbed'' by the inner chain resulting
in a larger inner chain (i.e., $N+1$ DSs) that simply oscillates in the
in-phase normal mode.
The mechanism whereby the outer DS is absorbed by the inner lattice
hinges on the fact that, although DS collision are elastic, there is
a shift in the path of the DSs with respect to the before and after collision
trajectories as was shown before (see discussion on the last collision
depicted in Fig.~\ref{fig:u2_collisions}).
The details of the outer DS ``absorption'', or rather the energy exchange
between the outer DS dynamics and the inner DS chain, is explained in
Fig.~\ref{fig:cradle_schematic}.
This energy transfer is more clearly visible in the left-bottom panel
of Fig.~\ref{fig:cradle_tf2000}
depicting the time derivative of the (square root of the) density. 
In this
panel, it is clear that during the first hypersoliton excursion through
the inner DS chain, there is practically no scattering of energy in the
inner chain. 
However, as explained in Fig.~\ref{fig:cradle_schematic},
just after the hypersoliton ejects the first DS, the inner chain has
an extra DS on the left and is missing one DS on the right and thus,
it is no longer close to equilibrium and starts to oscillate. This is
clearly visible in the bottom-left panel of Fig.~\ref{fig:cradle_tf2000}
for $300<t<500$ where the inner chain moves synchronously.
This energy transfer mechanism continues until all the energy of
the outer DS is completely depleted and the outer DS is 
absorbed by the inner lattice that oscillates with its
in-phase normal mode (results not shown here).
%

\begin{figure}[t] 
\begin{center}
\hskip-0.2cm
\psfig{figure=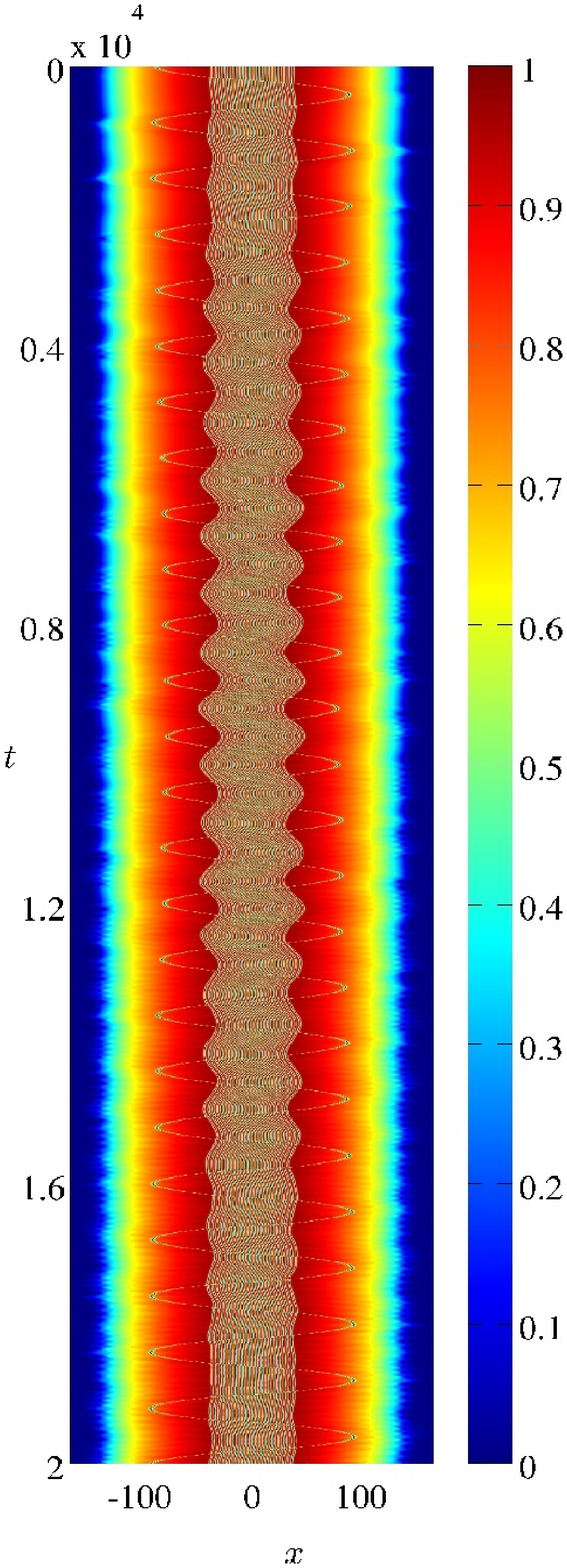,height=12.2cm,silent=}
~
\psfig{figure=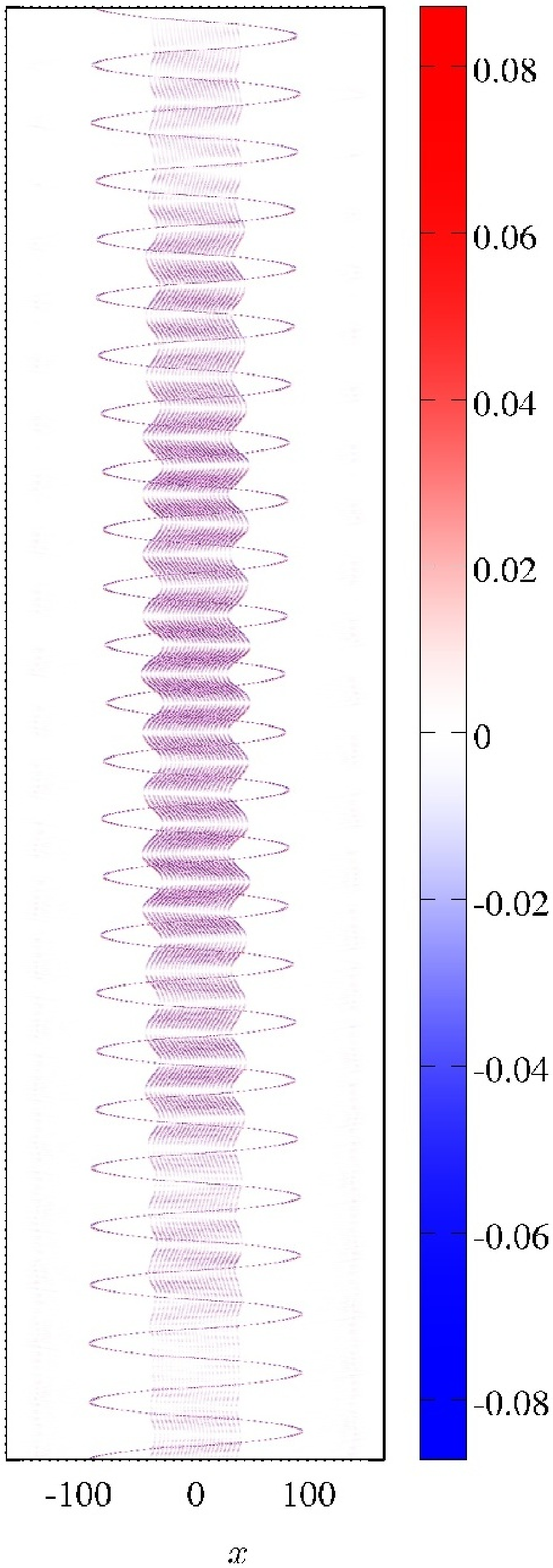,height=12.2cm,silent=}
\end{center}
\caption{{Color online}.
Long term evolution of the hypersoliton Newton's cradle corresponding
to the case depicted on the middle column of Fig.~\ref{fig:cradle_tf2000}.
Notice the beating of energy exchange between the outer DS and the
central chain.
All panels have the same meaning and layout as previous figures.}
\label{fig:cradle_tf20000}
\end{figure}

In order to avoid, or minimize, the energy transfer between the outer DS
and the inner DS chain, it is necessary to decouple the dynamics of the
outer DS and the inner DS chain. This is achieved by increasing the dropping
distance of the outer DS.
As can be seen in the case depicted in the middle column of panels in
Fig.~\ref{fig:cradle_tf2000},
the excursion of the outer DS has very little effect on the inner chain
motion. This can be explained because in this case, the outer DS travels
faster through the inner chain and thus, the latter has less time to develop
its in-phase normal mode. Furthermore, as the period of the
outer DS is different from the period of the inner chain in-phase mode,
subsequent excursions of the outer DS do not synchronize with the in-phase mode.
The outer DS has a different period in the presence of the inner
DS chain because its trajectory
in $(x,t)$ is the concatenation of sinusoids (when the outer DS is traveling
up and down the external trap on its own) and straight paths
(when the hypersoliton traverses the inner chain).\footnote{We note here that both the
sinusoid and straight paths are only approximate since (i) the outer DS
reaches regions relatively close to the edge of the condensate where the
approximation of harmonic motion of the DS is not longer accurate, and
(ii) the inner chain is not exactly equidistant and thus the hypersoliton
cannot travel strictly at a constant speed.}
Therefore, let us effectively consider the two involved dynamics, namely the outer soliton
oscillations and the inner chain oscillations, as two coupled oscillators.
These oscillators can synchronize provided that their periods are close to each
other and that the coupling is sufficiently strong~\cite{Strogatz:book}.
This is precisely what happens
for a relatively small dropping distance as depicted in the first case
of Fig.~\ref{fig:cradle_tf2000}. However, as we increase the dropping distance,
the two coupled oscillators have increasingly different frequencies and, at the
same time, the coupling is {\em reduced} as the interaction time between the
two, given by the time it takes for the hypersoliton to traverse the inner chain,
is reduced because of a faster hypersoliton speed.
Thus, in principle, there should be a threshold drop-off distance for which the
Newton's cradle should be self-maintained.
This is what seems to be occurring in the case depicted in the middle column of
Fig.~\ref{fig:cradle_tf2000} where apparently a very small amount of energy
is transferred to the inner chain.
To ensure that this small transfer does not destroy the Newton's cradle, we
depict in Fig.~\ref{fig:cradle_tf20000} the same case as in the middle column of
Fig.~\ref{fig:cradle_tf2000}, but for a longer time. As can be seen from the
figure, there is indeed a transfer of energy from the outer DS to the inner
chain for $t<9,000$. However, the roles are inverted after this time and then the
inner chain transfers back the energy to the outer DS! This produces a
beating phenomenon common for coupled oscillators with different
periods. This periodic energy transfer between the outer DS and the
inner chain seems to persists for very long times (results not shown here)
providing a mechanism for a stable, long-lived, Newton's cradle dynamics.

Finally, the last column of Fig.~\ref{fig:cradle_tf2000} depicts an example
with an even larger drop-off distance. In this case, the outer DS also interacts
with the edge of the BEC cloud and produces sounds waves that remove energy
from the former and thus, finally settling to a Newton's cradle with slightly
lower oscillation amplitude with some background radiation (sound waves)
prevailing in the condensate (results not shown here).

\section{Conclusions and outlook}
\label{sec:conclu}

We construct coherent structures consisting of compression
waves riding on chains of bright (BS) and dark (DS) solitons
of the GP model. Namely, a soliton riding on a chain of solitons
and thus dubbed a {\em hypersoliton}.
The dynamics for chains of BSs and DSs is reduced to a Toda lattice
on the solitons' positions, i.e., the solitons are modelled as
a chain of nonlinearly coupled masses. 
Then, the corresponding Toda lattice
solitons (compression waves on the lattice) can be initialized
on the original GP model using the exact Toda lattice soliton
solution.
We show how BS chains are inherently unstable due to phase desynchronization
between consecutive BSs and thus are poor candidates for supporting hypersolitons.
In contrast, DS chains are stable and DSs, being topologically charged, never
lose their phases and thus are always mutually repelled from each other.
We successfully craft hypersoliton solutions riding on DS chains of the original
GP model for a wide range of parameter values. These hypersolitons are
robust and stably travel at a constant speed without deformation
nor radiation.
Additionally, we construct multiple hypersolitons and observe their
elastic collisions in different head-on and chasing collisions scenarios.
Finally, inspired by the classical Newton's cradle, we study the dynamics
of finite DS chains trapped inside the customary parabolic external
potential relevant in experimental Bose-Einstein condensates.
This is achieved by letting a free, outer,
soliton hit a stationary inner DS chain creating a hypersoliton wave
travelling through the latter. As the hypersoliton reaches the end of
the inner chain, a single DS is expelled and allowed to rise and fall 
down the external trap hitting the inner chain repeating the
process in a manner akin to the classical Newton's cradle.
We study the effects of the drop-off distance on the formation of the
Newton's cradle dynamics and argue, in terms of the theory
of coupled oscillators, that a minimal drop-off distance
is required for the creation of self-sustained Newton's cradle
oscillations.

The present work could be extended in a few interesting directions.
For example the effects of finite temperatures in a condensate
give rise to dissipation due to coupling with the thermal
(non condensed) fraction. This dissipation can be modelled
at the level of the GP equation by the so-called phenomenological
dissipation~\cite{DJF_REVIEW,DARK_BOOK} and it is responsible for
anti-damping terms in the reduced equations of motion of the DSs.
It would be interesting to analyze the effects of such a dissipative
term on the dynamics of hypersolitons.
On the other hand, condensates can be supported by two or more
coupled components with linear and/or nonlinear coupling terms
between them~\cite{BEC_BOOK}. These coupled models give rise
to coupled complexes with dark or bright solitons coupled to
dark or bright solitons in the other component(s), thus giving
rise to the so called dark-dark and dark-bright
solitons~\cite{pe2,pe3,pe4,pe5}.
The dynamically reduced models for these coupled systems
take the form of coupled Toda
lattices~\cite{Gerdjikov_TL_Manakov}. It would be interesting to
explore the possibility to construct hypersolitons and study
their stability in systems with several components.

\bigskip

\section*{Acknowledgments}

M.M.~gratefully acknowledges support from the provincial Natural
Science Foundation of Zhejiang (LY15A010017) and the National
Natural Science Foundation of China (No.~11271342).
R.C.G.~gratefully acknowledges the support of NSF-DMS-1309035.

\bigskip

\end{document}